\documentclass{IET}

\usepackage{graphicx,subfigure}
\usepackage[T1]{fontenc}
\usepackage{algorithmic}
\usepackage{algorithm}
 \usepackage{lipsum}
\usepackage{mathtools}
\usepackage{cite}
 \usepackage{ragged2e}
\usepackage{multirow}
\usepackage{pbox}
\usepackage{array}
\usepackage{booktabs}

\usepackage{xcolor,colortbl}
\definecolor{gray}{rgb}{0.9, 0.9, 0.9}
\usepackage{amssymb}
\usepackage{textcomp}
\usepackage{float}
\usepackage{url}
\usepackage{hyperref}
\usepackage{amsmath}
\makeatletter
\renewcommand\paragraph{\@startsection{paragraph}{4}{\z@}%
            {-2.5ex\@plus -1ex \@minus -.25ex}%
            {1.25ex \@plus .25ex}%
            {\normalfont\normalsize\bfseries}}
\makeatother
\usepackage{titlesec}
\usepackage[english]{babel}
\usepackage{babel,blindtext}
\usepackage{pbox}
\usepackage{float}
\usepackage{marvosym}
\usepackage{booktabs}

\begin{document}

\title{Real-time Rescheduling in Distributed Railway Network: An Agent-Based Approach}

\author[1,\Letter]{Poulami Dalapati}
\affil{Dept of Information Technology, National Institute of Technology Durgapur, India}
\author[1]{Piyush Agarwal}
\author[1]{Animesh Dutta}
\author[2]{Swapan Bhattacharya}
\affil{Dept of Computer Science and Engineering, Jadavpur University, Kolkata, India}
\affil[\Letter]{E-mail: dalapati89@gmail.com}

\abstract{This paper addresses the issues concerning the rescheduling
of a static timetable in case of a disaster encountered in a large and
complex railway network system. The proposed approach tries to modify the schedule
so as to minimise the overall delay of trains. This is achieved by representing the
rescheduling problem in the form of a Petri-Net and the highly uncertain disaster
recovery times in such a model is handled as Markov Decision Processes $(MDP)$. For solving the 
rescheduling problem, a Distributed Constraint Optimisation $(DCOP)$ based strategy
involving the use of autonomous agents is used to generate the desired schedule.
The proposed approach is evaluated on the actual schedule of the Eastern Railways, 
India by constructing various disaster scenarios using the Java Agent
DEvelopment Framework $(JADE)$. When compared to the existing approaches, the proposed
framework substantially reduces the delay of trains after rescheduling.}

\maketitle
\section{Introduction}
\label{Introduction}
The railway system is a major mode of transport which is
geographically distributed throughout the country \cite{IR}. The construction of schedules for trains
in such system \cite{62, 64, 31, 32, 100} in an efficient and optimised manner is a challenging task with considerations
like situational complexity of the network and the enormous 
constraints that have to be handled. Some of these constraints
are availability of tracks between stations and availability of
platforms on those stations, which influence the arrival and departure
time of trains. Moreover, any disruption in railway network \cite{IR, accidents} due to 
natural calamities, sabotage, temporary platform blockage and accident
on track(s) or platform make the offline schedule sub-optimal for use.
Affected trains need to be rescheduled dynamically to minimise the impact
of such disruptions. Where not only the objective function changes over time but the constraints
can also transform \cite{1, fabien133}. This uncertainty and dynamic
constraints make the global optima less effective.
Therefore, the entire scheduling problem is considered as an agent based Distributed
constraint optimization problem (DCOP) \cite{33, 34}, where all agents
cooperate with each other using commonly agreed protocols and constraints.
The uncertainty of recovery time and its probabilistic nature is represented
mathematically in terms of Markov decision processes (MDP) \cite{124, 125, 127}. Here, each node
is considered as a possible state of disaster scenario in railway network.
The state transition functions are mapped to the constraints of DCOP, where
each agent chooses its action to minimise its expected delay based on
its policy. The action of the set of all agents in an MDP setting is to find
the optimal solution which minimises the total delay of the railway network
with the constraints in place. 
\\
\indent In this paper, a framework with multiple trains,
stations and tracks is considered where some of the trains are on tracks and
some others are at stations, as shown in Figure \ref{fig:railway}. In case of a disaster, in
and around a station, the station authorities inform the neighbouring stations,
the incoming and the outgoing trains. According to the proposed approach, each
train and station agent checks for disaster recovery time and resources available
to reach the destination. If the disaster recovery time does not affect the
scheduled arrival or departure time of trains, then original schedule is
maintained. Otherwise, the proposed rescheduling method, as described
in section \ref{Disaster_Handling_and_Rescheduling_Model},
is used to generate a new dynamic schedule and trains are informed. 
\\
\indent
The rest of the paper is organised as follows: In section \ref{Related_Work} some previous works
in related domain are discussed. Section \ref{Railway_Network} is devoted to the description of railway network
and scheduling topology. Section \ref{Railway_Architecture} models the system. 
DCOP and MDP representations of the system are depicted in section \ref{MDP_DCOP}.
Disaster handling and rescheduling approach is formulated in section \ref{Disaster_Handling_and_Rescheduling_Model}. The 
simulation results are evaluated in section \ref{Experiments_results}. Finally section \ref{Conclusion}
concludes the propose work with future scope.
\section{Literature Reviews}
\label{Related_Work}
The existing approaches mainly consider line topology and network topology while
rescheduling trains under disturbance \cite{84, 85, 86, ravi132}. The disturbance scenarios
are modelled as certain and uncertain \cite{61} based
on its recovery time. Further decision scenarios for 
rescheduling are classified as retiming, rerouting or reordering \cite{101, 104}
with respect to delay management of passenger railway services  \cite{85, 112, malik130}.\\
\indent 
A rescheduling system in tuberail trains dispatching problem is proposed in \cite{60}, 
which focuses on freight gross transport \cite{61} in time,
but not the delay time of passengers in railway and subway.
Here, the operation is centralised and
controlled by operation support system which communicates
with all trains, turnouts and handling equipment, so that all trains
are operated in global information condition.
A track-backup rescheduling approach \cite{62}
is proposed to minimise the negative effects
arising from the disturbances, which optimally assigns a backup track to each affected train,
based on original timetable, estimated recovery time, and track changing cost
in line topology. In \cite{116}, a heuristic-based mixed-integer linear programming model is proposed
to tackle delay propagation in traffic disturbances.
This model is robust to its configuration provided an appropriate selection of boosting method is performed.
A rescheduling model for last trains with the consideration of train delays caused by incidents that
occurred in train operations is also discussed in \cite{82}.
Here authors aim to minimise running-time, dwell-time (as defined in \cite{Dwell_time}), and differences between
rescheduled and original timetable and maximise average transfer redundant time. Similarly in \cite{101}, a proactive 
rerouting mechanism is proposed to minimise computational overhead
and congestion where links are affected by failures. An agent based game theoretic coalition formation model is proposed
in \cite{vito131} to re-optimise a railway timetable.
Train rerouting on an N-track railway network \cite{104} and robust railway station planning 
\cite{103} as well as optimisation in multi-train operation in subway system \cite{117} are also proposed to improve
the robustness of rescheduling process in the complex scenario.
\\
\indent
Although the train rescheduling problem has been widely studied, most of the previous work
consider either a centralised approach
or line topology. However, a real time railway network is distributed in nature with 
dynamic entities and disruption can happen anytime. Authorities
have to take decisions on the basis of 
real world scenario and concurrent decisions have to be made
for efficient handling of the problem.
Moreover, recovery time of such disturbances is highly uncertain \cite{30}. The
distributed nature of the system can be effectively represented
as DCOPs \cite{33, 34} and probabilistic decision making can be mapped with
MDP \cite{124}, where the outcomes are partly random and partly under the
control of decision maker. Again, the inherent dynamic nature and concurrent
decision making can be suitably modelled using Petri-Net \cite{119, 122, 120, 123, 129}.\\
\indent
In light of the discussion above, our \emph{main contributions} in this paper include:
\begin{itemize}
 \item Modelling of real-time railway system as a Petri-Net along with mathematical representation of
 the scenario with DCOP and MDP to enable formal analysis.
 \item An agent based disaster handling and rescheduling approach is also proposed
 considering network topology, which is capable of providing
 sufficiently good solution.
 \item Situational complexity of scalability issues in terms of number of
 decision variables and constraints are also taken into consideration
 while optimising total journey time delay of trains. 
\end{itemize}

\section{Railway Network and Scheduling Topology}
\label{Railway_Network}
A railway network consists of \emph{Stations ($S$)}, \emph{Trains ($T$)} and \emph{Tracks} as shown in Figure \ref{fig:railway}.
Multiple trains are either at stations or running on tracks at time instant $t$.
Railway scheduling and rescheduling is performed in such a way that, the highest priority train is rescheduled
first. Depending on the scenario, the priority changes dynamically reducing total journey time delay.
\subsection{Assumptions}
\begin{itemize}
 \item There can be multiple tracks between two stations.
 \item Each station may consist of multiple platforms.
 \item Stations can communicate with trains and neighbouring stations.
 \item Trains can communicate with stations only.
 \item Station conveys the recovery status of the blocked tracks or platform to trains and its neighbouring stations.
 \item All the trains begin and end their journey at stations.
\end{itemize}

\subsection{Classification of Railway Parameters}
\label{Classification}
\subsubsection{Stations $(S)$}
\begin{itemize}
 \item Stations where train $T$ is scheduled to stop.
 \item Stations where train $T$ does not stop, but station is a junction.
 \item Stations where train $T$ neither stops nor the station is a junction, but in case of inconvenience, train may stop, so that other trains can pass.
\end{itemize}
 So, for generalisation, dwell-time \cite{Dwell_time} is taken as a parameter. 
 If dwell-time of train $T$ at any stations $S$ is greater than zero, then $S$ is a stopping station for $T$.

\begin{figure}[H]
\centering
\subfigure[]{\label{fig:railway}\includegraphics[scale=0.4]{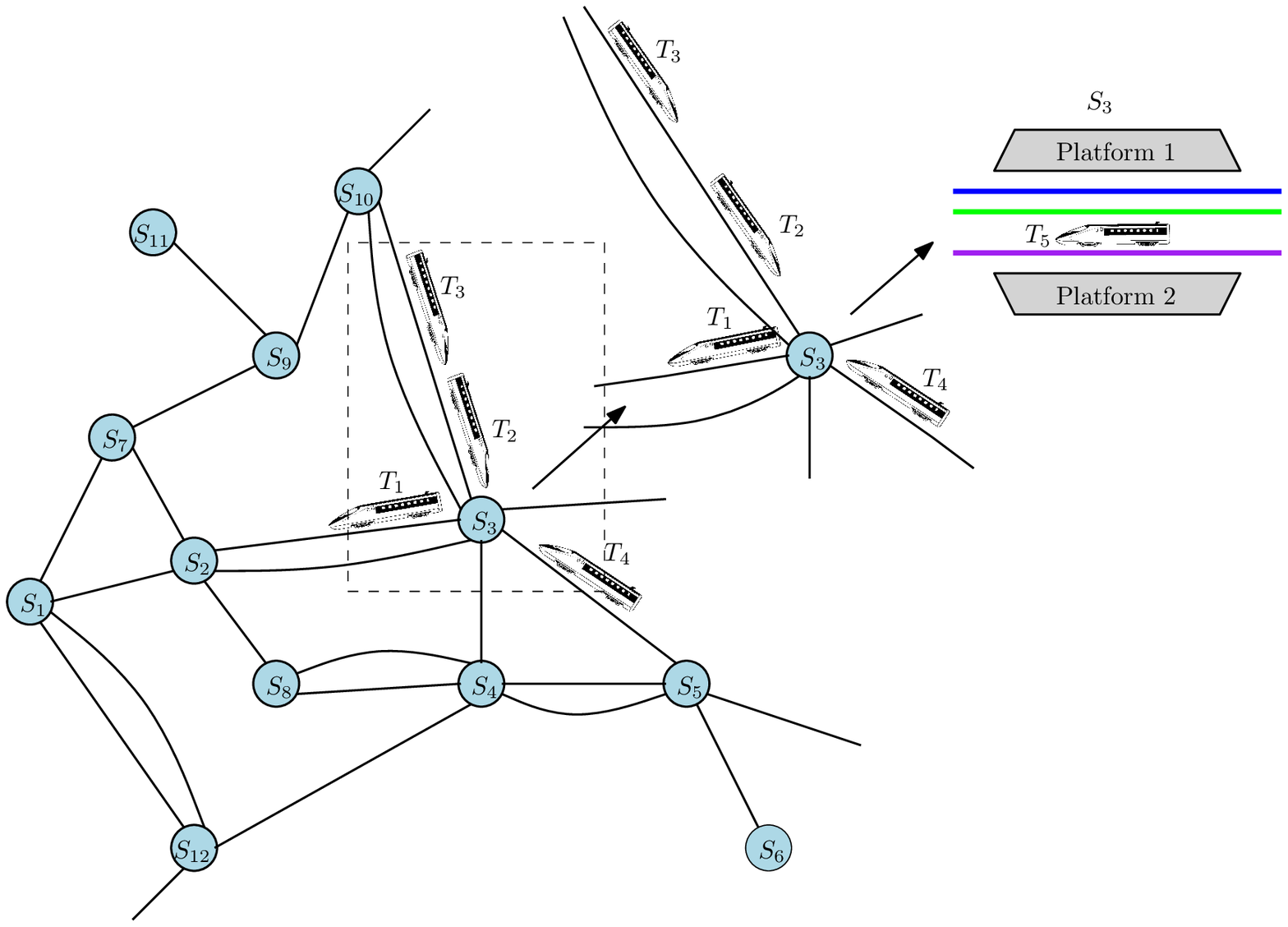}}
\hspace*{0.6cm}
\subfigure[]{\label{fig:platform}\includegraphics[scale=0.3]{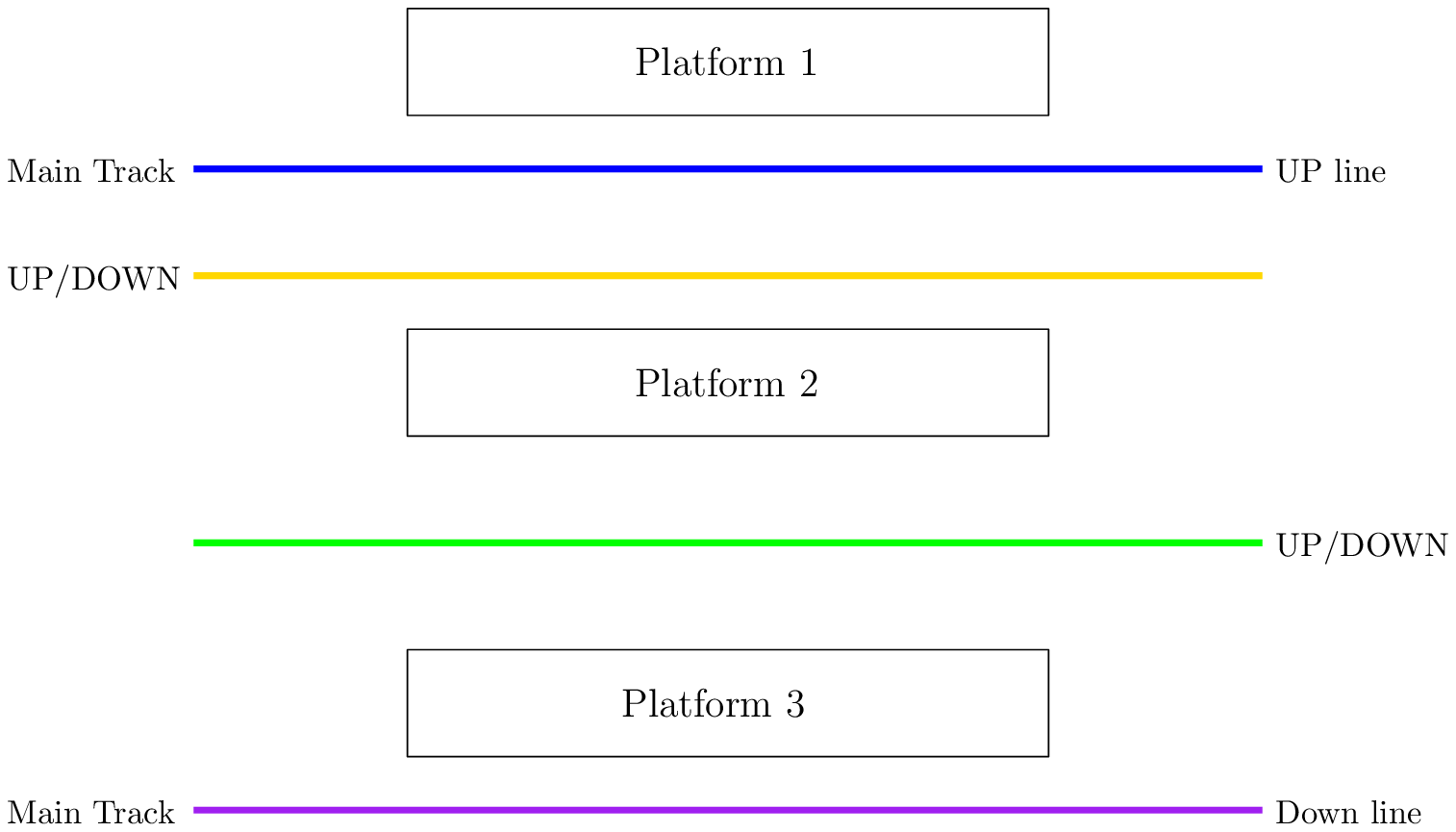}}
\caption{Railway Network.
\subcaption{a}{Graphical Representation of Railway Network.}
\subcaption{b}{Station with Multiple Tracks.}}
\end{figure}

\subsubsection{Tracks $(T)$ \newline}
As depicted in Figure \ref{fig:platform},
\begin{itemize}
 \item Double-line tracks between two stations, $UP$ and $DOWN$.
 \item Only a single line between two stations, used as either $UP$ or $DOWN$ as per schedule.
 \item Three or more tracks between two stations, $UP$, $DOWN$ and general tracks, used 
 as either $UP$ or $DOWN$ as per need.
\end{itemize}
For simplicity, we consider different directions of same track as two or more different tracks,
i.e. $UP$ as one resource and $DOWN$ as another, and 
rescheduling of either UP or DOWN line trains  at a time.

\subsubsection{Trains \newline}
\label{Prioritizing}

In a railway network \cite{IR},
depending upon various criteria, such as speed, facilities, distance covered, public demand, frequencies etc., train $T$ can be
 classified as Long-distance Train $(T^{L})$ and Short-distance Train $(T^S)$. $(T^{L})$ can again be categorised as
 Premium Train $(T^{Pr})$, Mail Trains $(T^{M})$, and Freight Train $(T^{F})$, whereas $(T^S)$ can be categorised as
 Passenger Train $(T^{P})$ and Local Train $(T^{Lo})$. 
 i.e., $T = T^{L} \cup T^{S}$, $T^{L} = T^{Pr} \cup T^{M} \cup T^{F}$, $T^{S} = T^{P} \cup T^{Lo}$.\\
\indent
Except $T^{Pr}$, during busy schedule (i.e. office hours), $T^S$
get higher priority than any $T^L$. Otherwise, during
normal hours, $T^L$ get higher priority. Again, as railway system faces delay due to many reasons,
these priorities change dynamically over time. As an example, if any $T^L$ is
delayed by more than the permissible threshold delay, then other trains get higher priority which are on time.
Priorities are assigned like, $y_{1}~=~Prio ~(T^{Pr})$,
$y_{2}~=~Prio ~(T^{M})$, $y_{3}~=~Prio ~(T^{F})$, $y_{4}~=~Prio ~(T^{P})$, $y_{5}~=~Prio ~(T^{Lo})$.
This priority allocation policy is defined by the existing railway system of the region, considered in the 
experiments as described in subsection \ref{Setup}.
In general, $y_{1} > y_{2} > y_{4} > y_{5} > y_{3}$. However when $t=t_{Busy}$
(from $9:00$ am to $11:00$ am and $5:00$ pm to $7:00$ pm), 
$y_{1} > y_{4} > y_{5} > y_{2} > y_{3}$. Again, in case of delayed trains, priority changes dynamically like,
$y_{4} \geq y_{1} \geq y_{5} \geq y_{2} > y_{3}$.

There are two kinds of inputs in our system, static input, which is pre-planned as per the schedule and
dynamic input, which is triggered by the changes due to disruption.
At any time instant $t$, each station has a fixed number of incoming and outgoing tracks. Station database
is updated with the information about incoming and outgoing trains in terms of their arrival and departure time. 
When a disaster occurs, one or more tracks
between stations get deleted from the databases and platform counts decreases 
from station databases. As railway network is represented as a connected multigraph, there may be
other possible paths to reach to the destination. 
After disaster has occurred, system checks for the trains which may reach a particular station within
the calculated buffer time $\tau^{B}$.

\section{Railway Architecture Model}
\label{Railway_Architecture}
Given this background, a \emph{Multi-agent System (MAS)} \cite{124, 128} is a natural choice for 
modelling such distributed system.
Here, we represent the railway network $(RN)$ as a pair of multigraph $(G)$ and an agency $(Ag)$.
i.e., $RN ~= ~<G, ~Ag>$.
Again, $G ~= ~<V, ~E>$, 
where $V$ is set of vertices and $E$ is set of edges.
\noindent From notations in Table \ref{Notations},
$V = \{v_i | ~i \in [1, n]\}$ and $v_i = S_i$ means vertex is a station, $E$ represents tracks between stations, 
$T = \{T_j | ~j \in [1, m]\}$, indicates trains (see Figure \ref{fig:railway}, and
$Ag = \{ Ag_a | ~a\in [1, q] \}$, denotes agency.
Each station and train is associated with an agent. $SA$ and $TA$ denote the set of station
agents and set of train agents respectively, where $S_i \in S$ with $Sa_a \in SA$ and $T_j\in T$
with $Ta_a \in TA$.
\begin{table}[!htbp]
 \caption{Notation}
\label{Notations}
\renewcommand{\arraystretch}{1.3}
\scalebox{0.70}{
\begin{tabular}{| l  p{9cm} || l  p{8cm} |}
\hline 
\rowcolor{gray}
\multicolumn{4}{|c|}{Indices and Parameters} \\
\hline \hline
$S$ & Stations & $o^{J}_{ji}$ & Journey time of train $j$ in original timetable\\
$T$ & Trains & $o^{d}_{ji}$ & Original dwell time of train $j$ at station $S_i$\\
$i$ & Station index & $x^{J}_{jl}$ & Journey time of train $j$ on track $l$ \\
$j$ & Train index & $a$ & Agent index\\
$l$ & Track index & $q$ & Number of agents, where $q=m+n$\\
$k$ & Platform index & $t$ & Time instant\\
$n$ & Number of stations & $t_D$ & Time of disaster\\
$m$ & Number of trains & $t_R$ & Time of recovery\\
$p$ & Maximum number of platforms at each station & $t_{Busy}$ & Busy Time Period of the day\\
$o^{AT}_{ji}$ & Arrival time of train $j$ at station $i$ in original timetable & $i' \in [1,n]\backslash i$ & Index of station other than the $i^{th}$ station.\\
$o^{DT}_{ji}$ & Departure time of train $j$ from station $i$ in original timetable & $j' \in [1,m]\backslash j$ & Index of train other than the $j^{th}$ train.\\
$\delta_{ji}$ & Delay of train $j$ at station $i$ & $i'' \in [1,n]\backslash i, i'$ & Index of station other than the $i^{th}$ and $i'^{th}$ station.\\
$\delta_{jl}$ & Delay of train $j$ on track $l$ & $j'' \in [1,m]\backslash j, j'$ & Index of train other than the $j^{th}$ and $j'^{th}$ train.\\
$\delta_{Th}$ & Threshold value for delay of all trains & $\tau^B$ & Buffer Time, where $t_{D}+\tau_{1} \leq \tau^{B} \leq t_{D}+\tau_{2}$\\
\hline \hline
\rowcolor{gray}
\multicolumn{4}{|c|}{Decision Variables} \\
\hline \hline
$x^{AT}_{ji}$ & Arrival time of train $j$ at station $i$ due to disaster & $x^{DT}_{ji}$ & Departure time of train $j$ from station $i$ due to disaster\\
$P_{jik}$ & Platform indicator, $P_{jik}=1$ if train $j$ occupies $k^{th}$ platform of station $i$, otherwise 0 & $L_{jil}$ & Track indicator, $L_{jil}=1$ if train $j$ occupies $l^{th}$ track connecting to station $i$, otherwise 0 and when $L_{jil}=1$, $L_{j'il}=0$ \\
$Prio(T_{j})$ & Priority of train $T_j$ & $x^{d}_{ji}$ & Actual operation time of train $j$ at station $S_i$\\
$\tau_{1}$ & Minimum time required to recover from the disaster & $\tau_{2}$ & Maximum time required to recover from the disaster \\
$\tau_R$ & Time to recover with the density function $\phi(x)$, where, $x\in[\tau_{1},\tau_{2}]$ & & \\
\hline
\end{tabular}
}
\end{table}

\subsection{Petri-Net Model of Railway System}
\begin{figure}[!htbp]
  \centering
\includegraphics[scale=0.450]{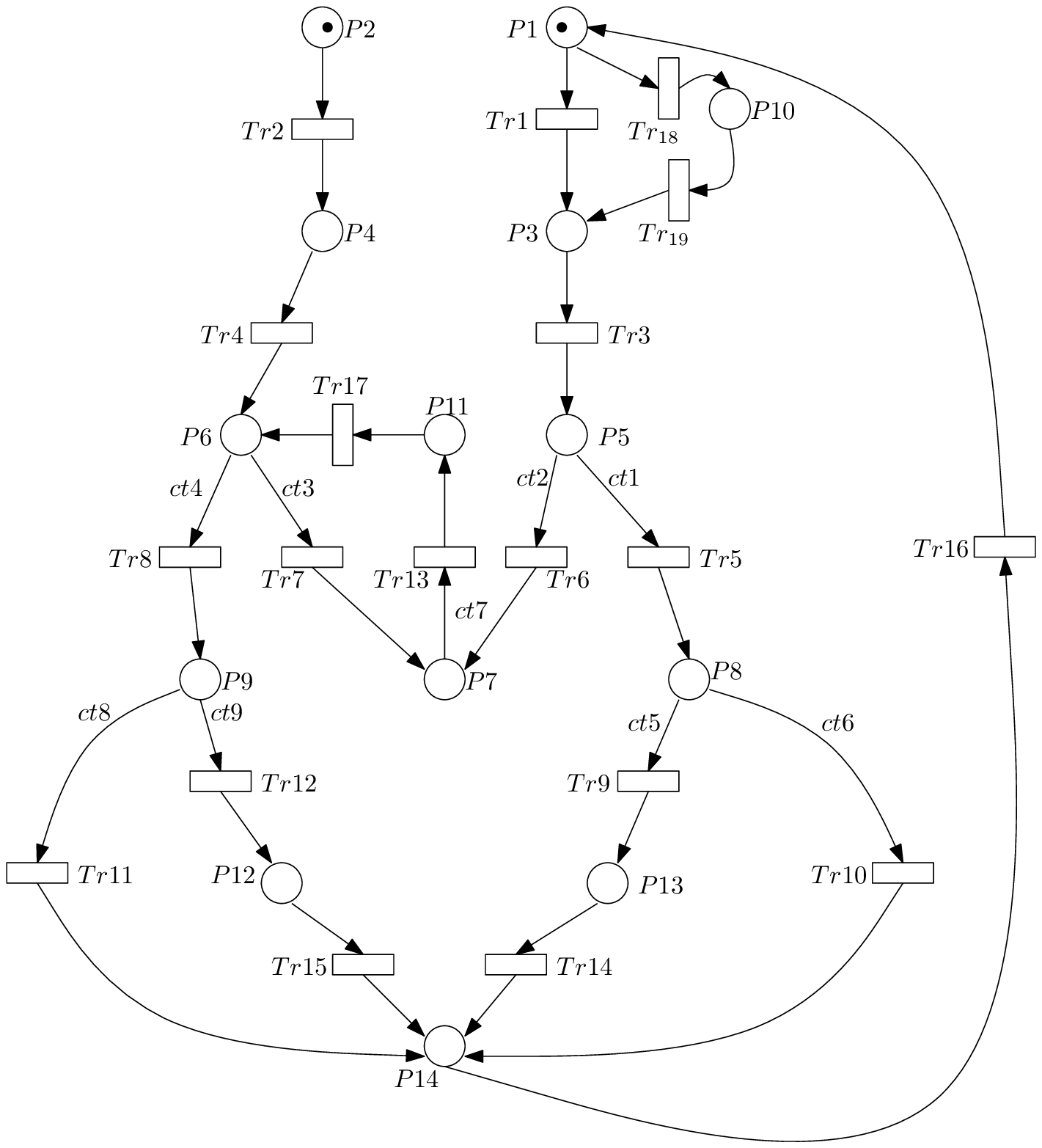}
\caption{Petri-Net model $PN1$ of Railway Network.}
\label{fig:PN1}
\end{figure}

Now we introduce the general concepts of \emph{Petri-Net} \cite{119, 122} describing a railway network. 
Major use of various kinds of Petri-Net \cite{120, 123, 129} is modelling of 
static and dynamic properties of complex systems, where concurrent
occurrences of events are possible, but there are constraints on the occurrences, precedence or frequency of these occurrences.
Graphically a Petri-Net is a directed bipartite graph where nodes represent places, transition and directed arcs which link
places to transitions or transitions to places. The state of a Petri-Net is given by the marking, describing the distribution of tokens in the places. 
Our proposed Petri-Net model deals with dynamism, uncertainty, and conflict situations 
in decision making choices upon different conditions. To overcome such conflicts the idea 
of colour token is introduced that enables a particular condition. 
Agents are considered as a token which can move from one
environmental state to other. In real time system agents perform some action if it sense
a particular environment; in contradiction some states are just used as an intermediate one.
The Petri-Net model for railway network is proposed as follows:\\
\\
\noindent
$\{P, Tr, F, Tok, f_{C}, M_{0}\}$ 
\\
\\
$P: \{P_{1}, P_{2}, \dots, P_{b}\}$, where $b>0$ is a finite set of \emph{Places}.\\
$P=P_{N} \cup P_{f_{c}}$, where $P_{N}$ is the set of places where no explicit function is executed on
arrival of resource token and $P_{f_{c}}$ is the set of places which executes a function or checks condition
on arrival of resource token.
\\
\\
$Tr: \{Tr_{1}, Tr_{2}, \dots, Tr_{z}\}$, where $z>0$ is a finite set of \emph{Transitions}.\\
$Tr=Tr_{I} \cup Tr_{c}$, where $Tr_{I}$ is the set of immediate transition which is fired as soon as
the required tokens are available at input place and an action is performed. $Tr_{c}$ is the set of colour transition
which is fired when the colour token is available in the input place.
\\
\\
$F:(P \times Tr) \cup (Tr \times P)$ is the set of \emph{Flow Function}.\\
$F=F_{+} \cup F_{-}$, where $F_{+}$ refers finite set of input flow and $F_{-}$ refers finite set of output flow.
\\
\\
$Tok:$ Set of \emph{Token}.\\
$Tok=Tok^{c} \cup Tok^{Ag}$ and $Tok^{c} \cap Tok^{Ag}=\phi$, where $Tok^{c}$ is the set of colour token, $c$ represents colour and 
$Tok^{Ag}$ is the resource token (Agent token).
\\
\\
$f_{c}: \{f_{c}1,f_{c}2, \dots, f_{c}u\}$, where $u\geq 0$ is the set of \emph{Functions}
that execute in $P_{f_{c}}$ when a resource token arrived at the place. Function can generate colour token or perform some operations.
\\
\\
$M_{0}:$ \emph{Initial Marking} of Petri-Net.
\\
\\
$\beta: F_{+}(Tok^{c} \times Tok^{Ag})\rightarrow F_{-}(Tok^{Ag})$ or $(Tok^{c} \times Tok^{Ag}) \rightarrow Tok^{Ag}$\\
\\
$\beta$ says that colour token is only used for taking a decision to resolve conflict. It won't propagate to next state.\\
\indent
In our Petri-Net model, shown in Figure \ref{fig:PN1}, place is represented by a circle, transition by a rectangle, 
and input and output flow by arrow and the corresponding description is given in Table \ref{tab:PN1}.

\begin{table}[!htbp]
\caption{Description of $PN1$}
\label{tab:PN1}

\centering
\renewcommand{\arraystretch}{1.3}
\scalebox{0.70}{
\begin{tabular}{|c||l|}
\hline 
\rowcolor{gray}
\multicolumn{2}{|c|}{Description of Places of $PN1$}\\
\hline \hline
\rowcolor{gray}
Places(P) & Description\\
\hline \hline
P1 & Trains $T_j$ is at station $S_i$ and count waiting time to leave.\\
\rowcolor{gray}
P2 & Trains $T_{j'}$ is at station $S_{i'}$ and count waiting time to leave.\\
P3 & $T_j$ starts running.\\
\rowcolor{gray}
P4 & $T_{j'}$ starts running.\\
P5 & $T_j$ is on track.\\
\rowcolor{gray}
P6 & $T_{j'}$ is on track.\\
P7 & $T_j$ and $T_{j'}$ both sense junction station and checks whether it has free platform.\\
\rowcolor{gray}
P8 & $T_j$ senses simple station and checks whether it has free platform.\\
P9 & $T_{j'}$ senses simple station and checks whether it has free platform.\\
\rowcolor{gray}
P10 & $T_j$ has completed its journey and ready for the next journey.\\ 
P11 & $T_j$ entering into station.\\ 
\rowcolor{gray}
P12 & No free platform for $T_{j'}$.\\
P13 & No free platform for $T_j$.\\ 
\rowcolor{gray}
P14 & $T_j$ and\/or $T_{j'}$ reached to the station.\\
\hline \hline
\rowcolor{gray}
\multicolumn{2}{|c|}{Description of Transitions of $PN1$}\\
\hline \hline
\rowcolor{gray}
Transitions(Tr) & Description\\
\hline \hline
Tr1 & It will fire when $T_j$ finishes its waiting time at $S_i$.\\
\rowcolor{gray}
Tr2 & It will fire when $T_{j'}$ finishes its waiting time at $S_i$.\\
Tr3 & It will fire when  $T_j$ is on track.\\
\rowcolor{gray}
Tr4 & It will fire when $T_{j'}$ is on track.\\
Tr5 & It will fire if $T_j$ senses a simple station.\\
\rowcolor{gray}
Tr6 & It will fire if $T_j$ senses a junction station.\\
Tr7 & It will fire if $T_{j'}$ senses a junction station.\\
\rowcolor{gray}
Tr8 & It will fire if $T_{j'}$ senses a simple station.\\
Tr9 & It will fire if no platform is free for $T_j$.\\
\rowcolor{gray}
Tr10 & It will fire when at least one platform is free for $T_j$.\\ 
Tr11 & It will fire when at least one platform is free for $T_{j'}$.\\ 
\rowcolor{gray}
Tr12 & It will fire if no platform is free for $T_{j'}$.\\
Tr13 & It will fire when at least one platform is free for highest priority train.\\
\rowcolor{gray}
Tr14 & It will fire if a platform gets free for $T_j$.\\
Tr15 & It will fire if a platform gets free for $T_{j'}$.\\
\rowcolor{gray}
Tr16 & It will fire when train entering into the station.\\
Tr17 & It will fire when train leaving a junction station.\\
\rowcolor{gray}
Tr18 & It will fire when $T_j$ reaches its destination.\\
Tr19 & It will fire when $T_j$ is ready to leave for a new journey.\\
\hline \hline
\rowcolor{gray}
\multicolumn{2}{|c|}{Description of colour tokens of $PN1$}\\
\hline \hline
\rowcolor{gray}
Colour Token(ct) & Description\\
\hline \hline
ct1 & P5 generates it if $T_j$ senses a simple station in front of it and enables transition Tr5.\\
\rowcolor{gray}
ct2 & P5 generates it if $T_j$ senses a junction station in front of it and enables transition Tr6.\\
ct3 & P6 generates it if $T_{j'}$ senses a junction station in front of it and enables transition Tr7.\\
\rowcolor{gray}
ct4 & P6 generates it if $T_{j'}$ senses a simple station in front of it and enables transition Tr8.\\
ct5 & P8 generates it if no free platform is available and enables transition Tr9.\\
\rowcolor{gray}
ct6 & P8 generates it if at least one platform is available and Tr10 is enabled  and enables transition Tr10.\\
ct7 & P7 generates it if at least one platform is available and enables transition Tr13.\\
\rowcolor{gray}
ct8 & P9 generates it if at least one platform is available and Tr11 is enabled and enables transition Tr11.\\
ct9 & P9 generates it if no free platform is available and enables transition Tr12.\\
\hline
\end{tabular}
}
\end{table}
\section{DCOP and MDP Representation of the Proposed Rescheduling Approach}
\label{MDP_DCOP}
\subsection{Representation as DCOP}
The problem of train rescheduling is represented as DCOP with four tuples, $\langle Ag, X, D, C  \rangle$, where, \\
\noindent
$Ag$ : Set of agents\\
$X$ : Set of variables, $X = \{x_{ji}^{AT} \cup x_{ji}^{DT} \cup P_{jik} \cup L_{jil}$ \}\\
$D$ : Set of domains, $D = t_D + \tau_R$\\
$C$ : Set of constraints
\subsubsection{Constraints $(C)$ \newline}
\noindent
$\bullet$ \textit{Continuity Constraint:}
 \begin{equation}
 \label{eq:Continuity}
  x^{AT}_{ji} \geq x^{DT}_{ji'} + x^{J}_{jl}
 \end{equation}
 i.e. the arrival time
 of any train at any station depends on its departure time from the previous station and the total journey time between these stations.\\
$\bullet$ \textit{Time Delay Constraint:}
 \begin{equation}
 \label{eq:Delay}
 \left.\begin{tabular}{l}
$x^{AT}_{ji}$ $\geq$ $o^{AT}_{ji}$\\
$x^{AT}_{ji}$ $-$ $o^{AT}_{ji}$ $= \delta_j$ 
\end{tabular}\right\}
 \end{equation}

i.e. the actual arrival time of $T_j$ at $S_i$ must be greater or equal to its original arrival time at that station.
If both are equal then the train is on time, otherwise there is some delay $\delta_j$. Delay is calculated by the difference between the original
journey time and the actual journey time.\\
$\bullet$ Index of platforms that trains occupy can not be greater than $p$. 
 \begin{equation}
 \label{eq:Platform}
  \forall j \in [1,m] ~\exists ~P_{jik} \in [0,1], ~where ~1\leq k \leq p, ~\forall ~S_{i}
 \end{equation}
 and
  \begin{equation}
  \label{eq:PlatformSum}
 \sum_k P_{jik} \leq p, ~where~ 1\leq k \leq p
 \end{equation}
 \noindent
$\bullet$ If train $T_{j}$ occupies $l^{th.}$ track, connecting to station $S_{i}$, then train $T_{j'}$
 can not occupy the same track at the same time.
  \begin{equation}
  \label{eq:TrackOccupy}
  if ~L_{jil}=1, ~then ~L_{j'il}=0
 \end{equation}
$\bullet$ Required resources of $T_{j}$ at time $t$, $Re(T_{j})|^{t}$, is either a platform at a station or a track between two stations.
 \begin{equation}
 \label{eq:Resources}
 Re(T_{j})|^{t} = L_{jil} ~or ~Re(T_{j})|^{t} = P_{jik}
 \end{equation}
$\bullet$ Route of the train $T_{j}$, $Rou(T_{j})$, is a series of $P$ and $L$.
 \begin{equation}
 \label{eq:Route}
   Rou(T_{j}) = (\bigwedge_{i=1}^{n-1} P_{jik}L_{jil}) \cup P_{jnk}, ~where ~j \in [1, ~m]
 \end{equation}
\subsection{Representation as MDP}
In real world, agents inhabit an environment whose state changes either
because of agent's action or due to some external event. Agents sense the state of world
and the choice of new state depends only on agent's
current state and agent's action.
\begin{itemize}
 \item \textbf{Set of World State $(W)$}\\
Here, $W$ represents the set of agent's state(s) in railway network under disturbance.
Train agents can sense three kind of states. If $S_i$ is assumed to be 
a station where disaster happens, then $T_j$ is either on track connecting $S_i$ 
or at a platform of $S_i$ or at a platform of station$S_{i'}$, connected to $S_i$.\\
i.e. $W=\{(L_{jil}=1), (P_{jik}=1), (P_{j'i'k}=1)\}$
\item \textbf{Transition Function $(\Psi)$}\\
 The  state transition function is denoted as $\Psi(\omega, C, \omega')$.
 In our proposed approach, each action $C$ maps to constraint(s) of DCOP
 to satisfy to reach from state $\omega$ to the next state $\omega'$, where $\omega, ~\omega' \in W$.
 
 If the train $T_j$ is on the $l^{th}$ track, connecting to station $S_i$, where disaster happened, i.e. $L_{jil}=1$,
 but no platform is available, i.e. $P_{jik}=1$, then $T_j$ must wait on the current track. So, there is no state change from the current state $L_{jil}=1$.
 Now, if platform is free, i.e. $P_{jik}=0$, then $T_j$ can reach to the next station. So, state transition from current state $L_{jil}=1$ is possible.
 
If $T_j$ is on $p^{th}$ platform of $S_i$ and the $l^{th}$ track is free but the platform at the next station $S_{i'}$
 is not free, i.e. $P_{ji'k'}=1$, no state change is possible from current state $P_{jik}=1$. Similarly, 
 even if there is free platform at $S_{i'}$,
 if the $l^{th}$ track is not free or both the track and platform is not available at time $t$, no state change is possible. 
 
 In Figure \ref{fig:MDP}, every node represents a state and each arc 
 represents an action which is indeed a constraint. The transition from one state to another state happens
iff the corresponding agent satisfies the specific constraint(s).

\begin{figure}[H]
\centering
\includegraphics[scale=0.45]{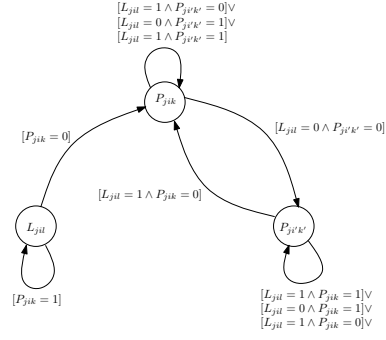}
\caption{MDP representation with set of states and transition functions.}
\label{fig:MDP}
\end{figure}
\end{itemize}

\section{Disaster Handling and Rescheduling Model}
\label{Disaster_Handling_and_Rescheduling_Model}
According to real-time scenario, in case of platform blockage
or track blockage, disaster handling and rescheduling model of railway system refers three situations:
\begin{itemize}
 \item \textit{Delay or Stop} at the station or on track $(\textbf{Retiming})$.
 \item \textit{Change in Departure Sequence} of trains at the station depending on priority of trains $(\textbf{Reordering})$.
 \item \textit{Reschedule} to alternative path $(\textbf{Rerouting})$.
\end{itemize}
\subsection{\textbf{Case 1: Partial Node Deletion from the graph $G$}}
\label{Case1}
A station $S_i$ faces problem due to disaster and the train $T_j$ is on track $l$, approaching to the station $S_i$,\\ i.e.
\begin{equation}
\label{eq:Case1}
L_{jil}=1 
\end{equation}
\subsubsection{Case 1.1 \newline }
If $S_i$ has a free platform at the time when $T_j$ reaches to $S_i$,
 then the system can allow $T_j$ to reach $S_i$, iff the priority of the incoming train $T_j$ has the highest priority
 amongst all trains $T$ and the resource $(Re)$ required for any other high priority train $T_{j''}$ does not hamper the resource requirement of $T_j$.
 \begin{equation}
 \label{eq:Case1.1}
  P_{jik}|_{x_{ji}^{AT}} = 1, ~ iff ~[Prio(T_{j}) > Prio(T_{j'})_{j \neq j',~ j\in [1,~m]}] ~\wedge [Re(T_{j''})|^{\tau^{B}}_{Prio(T_{j''}) > Prio(T_{j})} \neq Re(T_{j})|^{\tau^{B}}]
 \end{equation}
\paragraph{Case 1.1.1}
After satisfying the condition described in case 1.1,
if all the necessary resources are available throughout its journey, selecting
   any alternative path, then reroute train $T_j$. Route of $T_{j}$ after departure from $S_{i}$ at time $x_{ji}^{DT}$
   is $[Rou(T_{j})|_{t \leq x_{ji}^{DT}} + Rou(T_{j})|_{x_{ji}^{AT} > x_{ji}^{DT}}]$.\\
   i.e. 
   \begin{equation}
   \label{eq:Case1.1.1}
   if~ Rou(T_{j})|_{x_{ji}^{AT} > x_{ji}^{DT}} = \bigwedge_{i' > i}^{n-1} (P_{ji'k}L_{ji'l}) \cup P_{jnk} = 0
   \end{equation}
   then reschedule $T_j$.
\paragraph{Case 1.1.2}
If case 1.1.1 is invalid, then
stop $T_{j}$ at $S_{i}$ until recovery is done or any other alternative path
   becomes free. So, $T_j$ occupies one of the platforms at $S_i$.\\ i.e.
   \begin{equation}
   \label{eq:Case1.1.2a}
    P_{jik}=1
    \end{equation}
    and
    \begin{equation}
     \label{eq:Case1.1.2b}
    {x_{ji}^{DT}} = {o_{ji}^{DT}} + \delta_j
   \end{equation}
   where,
   \begin{equation*}
    \label{eq:Case1.1.2Delta}
    \delta_j = t_R - o_{ji}^{DT} ~and ~t_R = t_D + \tau_R
   \end{equation*}
 \subsubsection{Case 1.2 \newline }
\begin{figure}[H]
\centering
\subfigure[]{\label{fig:PN2}\includegraphics[scale=0.3]{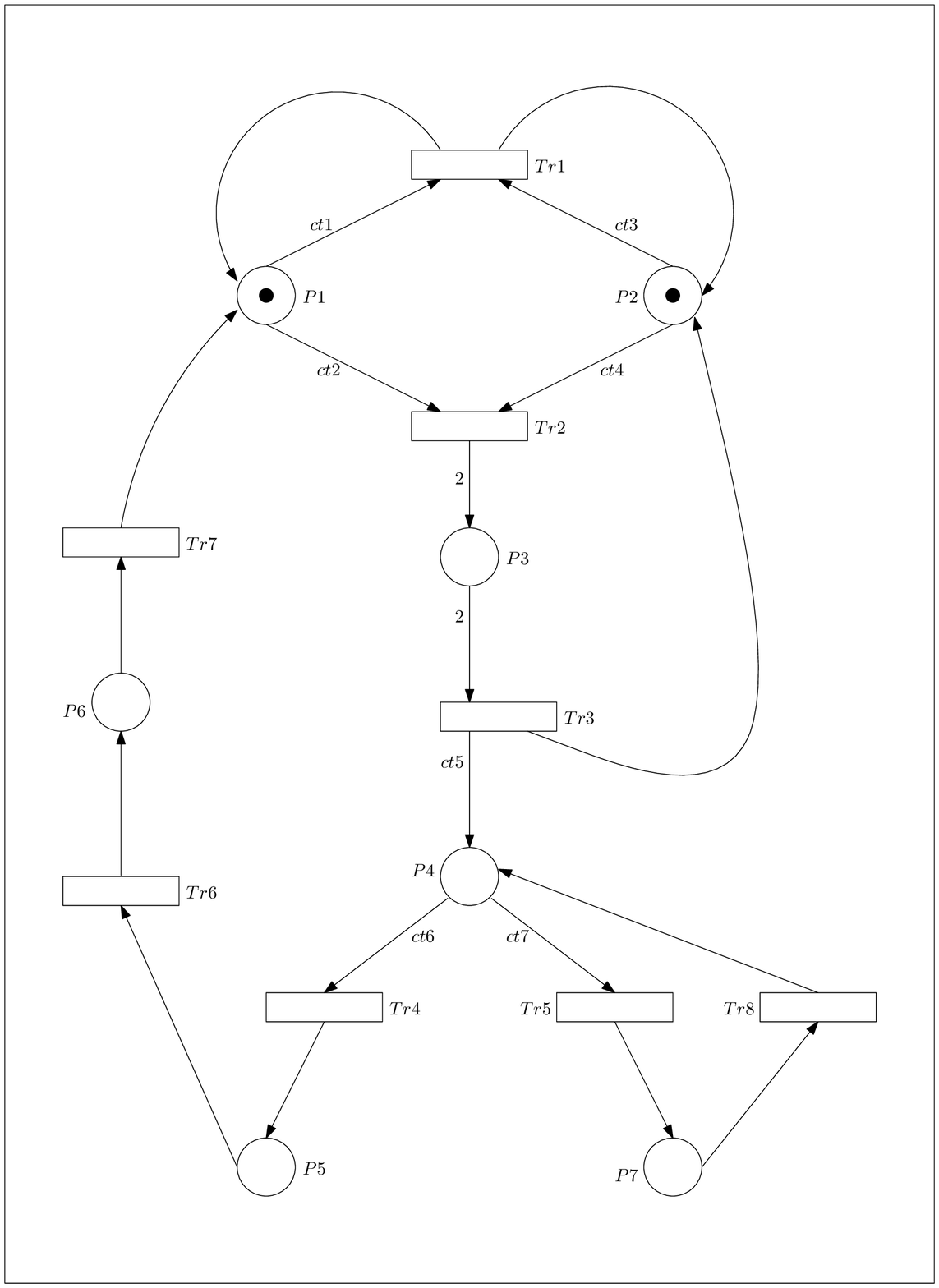}}
 \hspace{1.5cm}
\subfigure[]{\label{fig:PN2_seq}\includegraphics[scale=0.5]{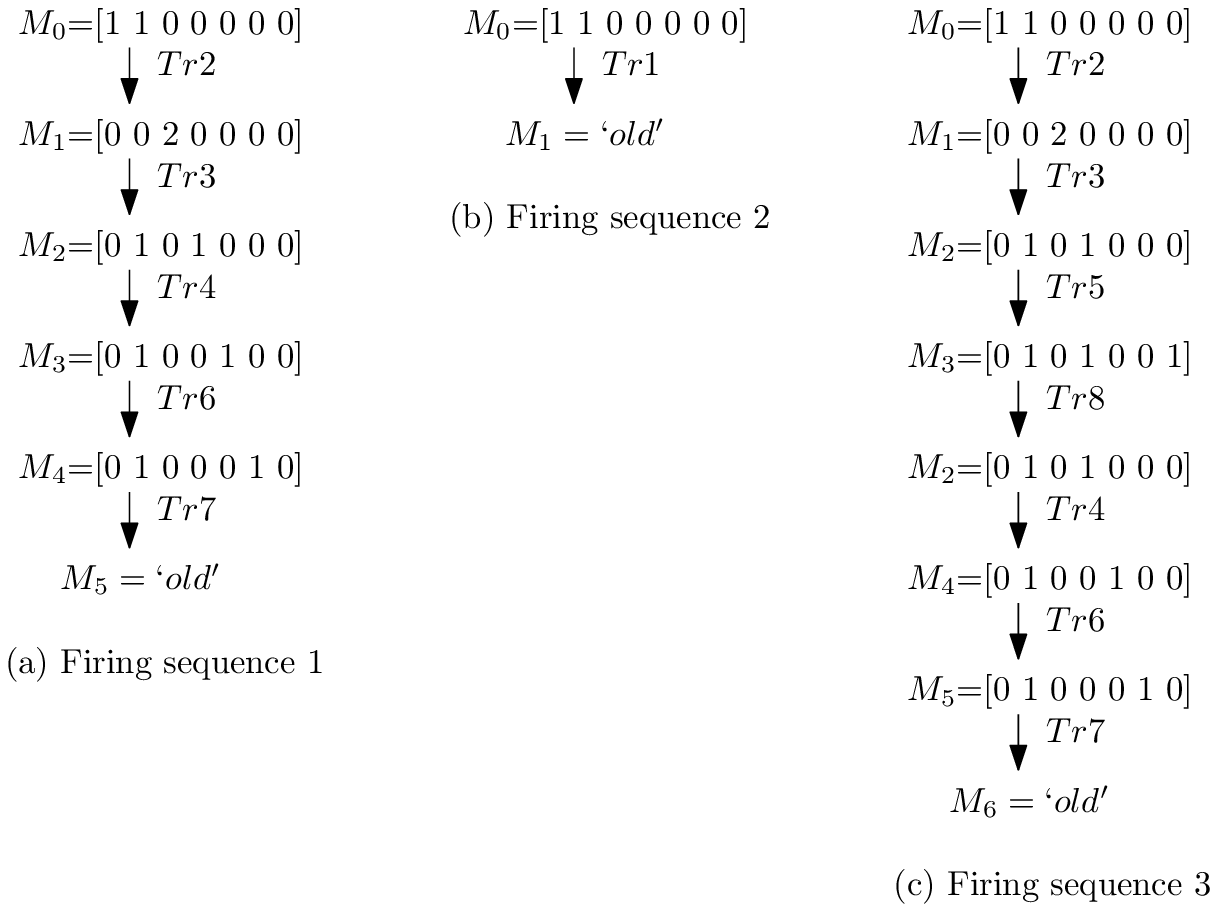}}
\caption{Petri-Net $PN2$
\subcaption{a}{Petri-Net model $PN2$ of Case 1.}
\subcaption{b}{Reachability tree of $PN2$ for different firing sequences.}}
\end{figure}

  If the scenario does not conform with case 1.1, then
  stop train $T_j$ on the current track. So, now $T_j$ occupies $l^{th}$ track.\\
  i.e.
  \begin{equation}
   \label{eq:Case1.2a}
   L_{jil}=1
  \end{equation}
 and
 \begin{equation}
  \label{eq:Case1.2b}
  {x_{ji}^{AT}} = {o_{ji}^{AT} + \delta_j}
 \end{equation}
The above described scenario depicted in equations (\ref{eq:Case1})-(\ref{eq:Case1.2b}) is now represented in Petri-Net model
in Figure \ref{fig:PN2} and the corresponding
 description of respective places, transitions and tokens are described in Table \ref{tab:PN2}.

\begin{table}[!htb]
\caption{Description of $PN2$}
\label{tab:PN2}

\centering
\renewcommand{\arraystretch}{1.3}
\scalebox{0.70}{
\begin{tabular}{|c||l|}
\hline 
\rowcolor{gray}
\multicolumn{2}{|c|}{Description of Places of $PN2$}\\
\hline \hline
\rowcolor{gray}
Places(P) & Description\\
\hline \hline
P1 & $T_j$ is on track $l$ and approaching to disastrous station $S_i$.\\
\rowcolor{gray}
P2 & $T_{j'}$ is on track $l'$ and approaching to disastrous station $S_i$.\\
P3 & $T_j$ and $T_{j'}$ both approaching to same station $S_i$.\\
\rowcolor{gray}
P4 & $T_j$ reaches to $S_i$ and is at $S_i$.\\
P5 & $T_j$ reaches its destination.\\
\rowcolor{gray}
P6 & $T_j$ completed its previous journey and started the next.\\
P7 & $T_j$ is waiting for availability of resources.\\
\hline \hline
\rowcolor{gray}
\multicolumn{2}{|c|}{Description of Transitions of $PN2$}\\
\hline \hline
\rowcolor{gray}
Transitions(Tr) & Description\\
\hline \hline
Tr1 & It will fire if platform is not available at $S_i$ the time of Arrival.\\
\rowcolor{gray}
Tr2 & It will fire if platform is available at $S_i$ the time of Arrival.\\
Tr3 & It will fire if $T_j$ gets highest priority.\\
\rowcolor{gray}
Tr4 & It will fire if all resources are available to Continue the journey.\\
Tr5 & It will fire if all resources are not available to Continue the journey.\\
\rowcolor{gray}
Tr6 & It will fire when $T_j$ reaches its destination.\\
Tr7 & It will fire when $T_j$ is ready for its new journey.\\
\rowcolor{gray}
Tr8 & It will fire when $T_j$ is not allowed to start its journey.\\
\hline \hline
\rowcolor{gray}
\multicolumn{2}{|c|}{Description of colour tokens of $PN2$}\\
\hline \hline
\rowcolor{gray}
Colour Token(ct) & Description\\
\hline \hline
ct1 & P1 generates it when equation (\ref{eq:Case1}) is satisfied but $T_j$ senses no free platform is available and enables transition Tr1. \\
\rowcolor{gray}
ct2 & P1 generates it when equation (\ref{eq:Case1}) is satisfied but $T_j$ senses free platform is available and enables transition Tr2.\\
ct3 & P2 generates it when equation (\ref{eq:Case1}) is satisfied but $T_{j'}$ senses no free platform is available and enables transition Tr1. \\
\rowcolor{gray}
ct4 & P2 generates it when equation (\ref{eq:Case1}) is satisfied but $T_{j'}$ senses free platform is available and enables transition Tr2.\\
ct5 & P3 generates it to denote $T_j$ has highest priority and enables transition Tr3.\\
\rowcolor{gray}
ct6 & P4 generates it $T_j$ senses all resources are available and enables transition Tr4.\\
ct7 & P4 generates it $T_j$ senses all resources are available and enables transition Tr5.\\
\hline
\end{tabular}
}
\end{table}
\noindent
\emph{\underline{Analysis of $PN2$:}}\\
Table \ref{tab:PN2} presents the description of
the places $P=\{P1, P2, P3, P4, P5, P6, P7\}$ and transitions $Tr=\{Tr1, Tr2, Tr3, Tr4, Tr5, Tr6, Tr7, Tr8\}$ and
the initial marking is $M_0=[1, 1, 0, 0, 0, 0, 0]$.
\begin{itemize}
\label{description:Analysis}
 \item Reachability graph analysis:\\
Reachability graph analysis is the simplest method to analyse
the behaviour of a Petri-Net. It decides whether the system
is bounded and live or not. From our
resultant tree in \ref{fig:PN2_seq} it can be proved that: a) the reachability set
$R(M_0)$ is finite, b)
maximum number of tokens that a place can have is 2, so our
$PN2$ is 2-bounded, c) all transitions can be fired, so there are
no dead transitions.
 \item State equation:\\
 The structural behaviour of the Petri-Net can be measured by
using the algebraic analysis of the incidence matrix. If marking $M$
is reachable from initial marking $M_0$ through the transition
sequence $\sigma$, then the following state equation holds:
$\boldsymbol M_0 + [\boldsymbol A] \times X_\sigma = \boldsymbol M$.\\
Incidence matrix is defined as 
$\boldsymbol{A}$ = [$e_{uv}$], it is a $r_A \times c_A $ matrix where
 ($1\leq u \leq r_A$), ($1 \leq v \leq c_A$). The order of the places in the
matrix is $P = \{P1, P2, P3, P4, P5, P6, P7\}$, denoted by rows 
and the order of the transitions is $Tr=\{Tr1, Tr2, Tr3, Tr4, Tr5, Tr6, Tr7, Tr8\}$, denoted by columns.\\
$X_{\sigma}$ is an m-dimensional vector with its ${j}^{th}$ entry denoting the number of
times transition $t_{j}$ occurs in $\sigma$. 
\[\tiny
\boldsymbol A = 
\begin{bmatrix}
    0 & -1 & 0 & 0 & 0 & 0 & 1 & 0 \\
    0 & -1 & 1 & 0 & 0 & 0 & 0 & 0 \\
    0 & 1 & -1 & 0 & 0 & 0 & 0 & 0 \\
    0 & 0 & 1 & -1 & -1 & 0 & 0 & 1 \\
    0 & 0 & 0 & 1 & 0 & -1 & 0 & 0 \\
    0 & 0 & 0 & 0 & 0 & 1 & -1 & 0 \\
    0 & 0 & 0 & 0 & 1 & 0 & 0 & -1 \\
\end{bmatrix}
\]

Thus, if we view a marking $\boldsymbol M_0$ as a k-dimensional
column vector in which the ${i}^{th}$ component is $M_0(p_i)$, each column of $[\boldsymbol A]$ is then
a k-dimensional vector such that  $\boldsymbol M_0 \xrightarrow \sigma \boldsymbol M $.\\

In our system, marking $\boldsymbol M=[1, 1, 0, 0, 0, 0, 0]$ is reachable from initial marking
$ \boldsymbol M_0=[1, 1, 0, 0, 0, 0, 0]$ through the firing sequence $\sigma_1 = Tr2, Tr3, Tr4, Tr6, Tr7$.\\ 
$M_0 \xrightarrow{Tr2} M_1 \xrightarrow{Tr3} M_2 \xrightarrow{Tr4} M_3 \xrightarrow{Tr6} M_4 \xrightarrow{Tr7} M_5(=\textquotesingle old \textquoteright)$.\\

 \[\tiny
\begin{bmatrix}
     1 \\
     1 \\
     0 \\
     0 \\
     0 \\
     0 \\
     0
\end{bmatrix}
+
\begin{bmatrix}
    0 & -1 & 0 & 0 & 0 & 0 & 1 & 0 \\
    0 & -1 & 1 & 0 & 0 & 0 & 0 & 0 \\
    0 & 1 & -1 & 0 & 0 & 0 & 0 & 0 \\
    0 & 0 & 1 & -1 & -1 & 0 & 0 & 1 \\
    0 & 0 & 0 & 1 & 0 & -1 & 0 & 0 \\
    0 & 0 & 0 & 0 & 0 & 1 & -1 & 0 \\
    0 & 0 & 0 & 0 & 1 & 0 & 0 & -1 \\
\end{bmatrix}
\times
\begin{bmatrix}
     0 \\
     1 \\
     1 \\
     1 \\
     0 \\
     1 \\
     1 \\
     0
\end{bmatrix}
=
\begin{bmatrix}
     1 \\
     1 \\
     0 \\
     0 \\
     0 \\
     0 \\
     0
\end{bmatrix}
\]

Similarly, marking $\boldsymbol M=[1, 1, 0, 0, 0, 0, 0]$ is reachable from initial marking
$\boldsymbol M_0=[1, 1, 0, \newline 0, 0, 0, 0]$ through the firing sequence $\sigma_2 = Tr1$ and $\sigma_3 = Tr2, Tr3, Tr5, Tr8, Tr4, Tr6, Tr7$. \\
$M_0 \xrightarrow{Tr2} M_1(=\textquotesingle old \textquoteright)$.\\
$M_0 \xrightarrow{Tr2} M_1 \xrightarrow{Tr3} M_2 \xrightarrow{Tr5} M_3 \xrightarrow{Tr8} M_2 \xrightarrow{Tr4} M_4 \xrightarrow{Tr6} M_5
\xrightarrow{Tr7} M_6(=\textquotesingle old \textquoteright)$.\\

\[\tiny
\begin{bmatrix}
     1 \\
     1 \\
     0 \\
     0 \\
     0 \\
     0 \\
     0
\end{bmatrix}
+
\begin{bmatrix}
    0 & -1 & 0 & 0 & 0 & 0 & 1 & 0 \\
    0 & -1 & 1 & 0 & 0 & 0 & 0 & 0 \\
    0 & 1 & -1 & 0 & 0 & 0 & 0 & 0 \\
    0 & 0 & 1 & -1 & -1 & 0 & 0 & 1 \\
    0 & 0 & 0 & 1 & 0 & -1 & 0 & 0 \\
    0 & 0 & 0 & 0 & 0 & 1 & -1 & 0 \\
    0 & 0 & 0 & 0 & 1 & 0 & 0 & -1 \\
\end{bmatrix}
\times
\begin{bmatrix}
     1 \\
     0 \\
     0 \\
     0 \\
     0 \\
     0 \\
     0 \\
     0
\end{bmatrix}
=
\begin{bmatrix}
     1 \\
     1 \\
     0 \\
     0 \\
     0 \\
     0 \\
     0
\end{bmatrix}
\]

\[\tiny
\begin{bmatrix}
     1 \\
     1 \\
     0 \\
     0 \\
     0 \\
     0 \\
     0
\end{bmatrix}
+
\begin{bmatrix}
    0 & -1 & 0 & 0 & 0 & 0 & 1 & 0 \\
    0 & -1 & 1 & 0 & 0 & 0 & 0 & 0 \\
    0 & 1 & -1 & 0 & 0 & 0 & 0 & 0 \\
    0 & 0 & 1 & -1 & -1 & 0 & 0 & 1 \\
    0 & 0 & 0 & 1 & 0 & -1 & 0 & 0 \\
    0 & 0 & 0 & 0 & 0 & 1 & -1 & 0 \\
    0 & 0 & 0 & 0 & 1 & 0 & 0 & -1 \\
\end{bmatrix}
\times
\begin{bmatrix}
     0 \\
     1 \\
     1 \\
     1 \\
     1 \\
     1 \\
     1 \\
     1
\end{bmatrix}
=
\begin{bmatrix}
     1 \\
     1 \\
     0 \\
     0 \\
     0 \\
     0 \\
     0
\end{bmatrix}
\]
\end{itemize}

\subsection{\textbf{Case 2: Partial Node Deletion from the graph $G$ with Deletion of an Edge}}
\label{Case2}
Currently multiple trains are on various platforms which follow a sequence according to their departure time within the buffer time $\tau^{B}$.
Here, $\tau^{B}$ is related to the disaster recovery time $t_R$ of that station.
  The track is free but more than one trains are yet to come.\\
  i.e.
  \begin{equation}
   \label{eq:Case2a}
  L_{jil}=0 
  \end{equation}
  and
\begin{equation}
 \label{eq:Case2b}
 \sum_{j=1}^{m} P_{jik} \leq (p-1)
\end{equation}
 As the station $S_i$ faces disaster, a particular $k^{th}$ platform can not be used until the recovery time has elapsed.
  If there is any incoming train $T_{j}$ within buffer period $\tau^{B}$, the system allows $T_{j}$ to reach $S_{i}$
  if a platform is available, i.e. $P_{jik}=0$. The system also checks for the priority of $T_j$ to reorder the departure schedule of all trains from $S_i$ introducing
  delay $\delta_{ji}$ to $T_j$, if needed.\\
  i.e. $\forall j$, if $Prio(T_{j'}) > Prio(T_{j})$
  \begin{equation}
  \label{eq:Case2c}
  {x_{ji}^{DT}} = {o_{ji}^{DT}} + \delta_{ji} 
  \end{equation} 
  
\begin{figure}[H]
\centering
\subfigure[]{\label{fig:PN3}\includegraphics[scale=0.70]{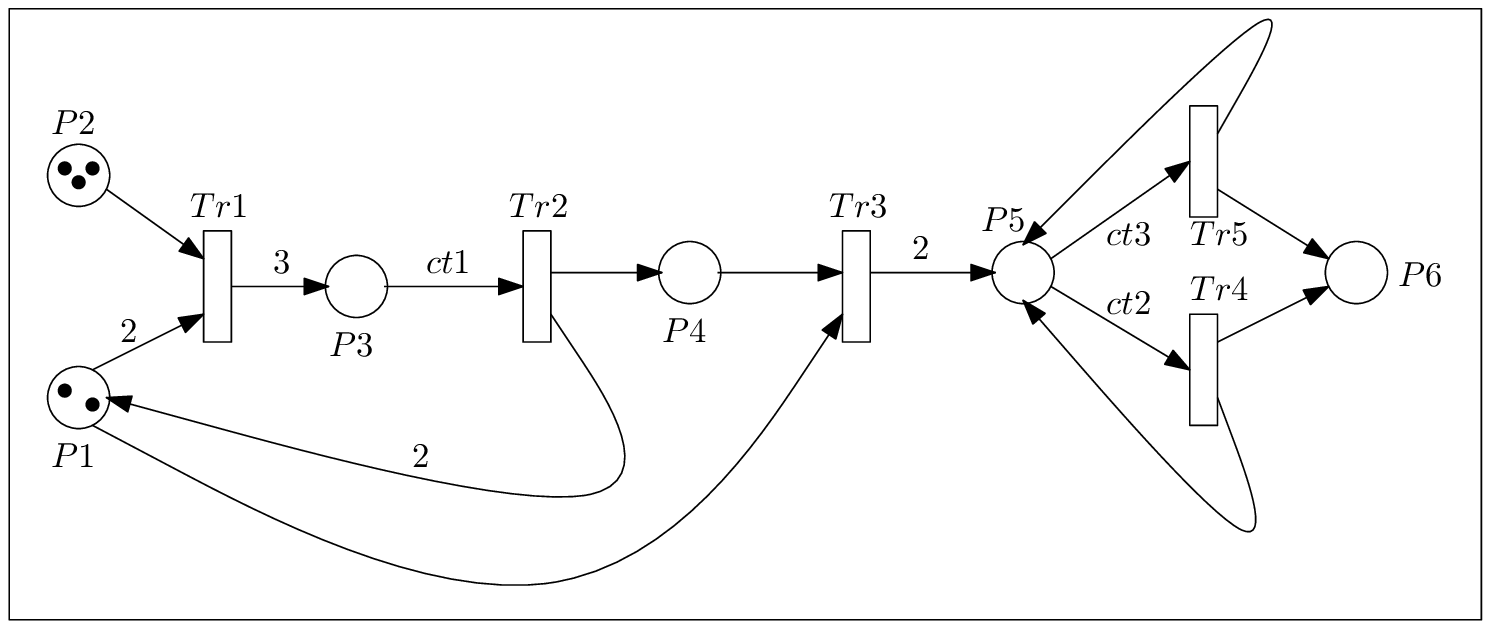}}
\hspace{1.5cm}
\subfigure[]{\label{fig:PN3_seq}\includegraphics[scale=0.80]{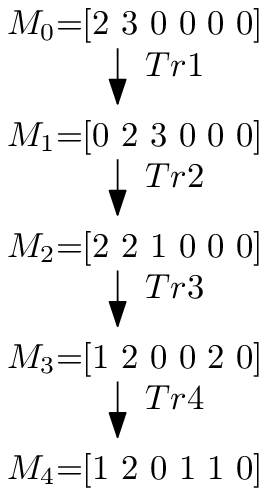}}
\caption{Petri-Net $PN3$
\subcaption{a}{Petri-Net model $PN3$ of Case 2.}
\subcaption{b}{Reachability tree of $PN3$ for different firing sequences.}}
\end{figure}

  Otherwise, if all the resources are available for $T_{j}$ and it has the highest priority among all the trains currently waiting at $S_{i}$,
  the scheduled departure of $T_{j}$ is the original departure time as per the original railway timetable.\\
  i.e.
  \begin{equation}
  \label{eq:Case2d}
   {x_{ji}^{DT}} = {o_{ji}^{DT}}
  \end{equation}
iff $Prio(T_{j}) > Prio(T_{j'})_{j \neq j',~ j\in [1,~m]}$\\
\\
The scenario of Case 2 with equations (\ref{eq:Case2a})-(\ref{eq:Case2d}) is represented in Petri-Net model in Figure \ref{fig:PN3} and the corresponding
 description of respective places, transitions and tokens are described in Table \ref{tab:PN3}.

\begin{table}[H]
\caption{Description of $PN3$}
\label{tab:PN3}

\centering
\renewcommand{\arraystretch}{1.3}
\scalebox{0.70}{
\begin{tabular}{|c||l|}
\hline 
\rowcolor{gray}
\multicolumn{2}{|c|}{Description of Places of $PN3$}\\
\hline \hline
\rowcolor{gray}
Places(P) & Description\\
\hline \hline
P1 & Trains are at $S_i$ where disaster happens and only one platform is free.\\
\rowcolor{gray}
P2 & Trains are at station connecting to $S_i$.\\
P3 & More than one trains are requesting for a single platform.\\
\rowcolor{gray}
P4 & Highest priority train reaches to $S_i$.\\
P5 & All the trains are at $S_i$ waiting to depart.\\
\rowcolor{gray}
P6 & Highest priority train departs from $S_i$.\\
\hline \hline
\rowcolor{gray}
\multicolumn{2}{|c|}{Description of Transitions of $PN3$}\\
\hline \hline
\rowcolor{gray}
Transitions(Tr) & Description\\
\hline \hline
Tr1 & Only one platform is available at $S_i$ and more than one train are approaching to $S_i$.\\
\rowcolor{gray}
Tr2 & One platform is available and $T_j$ has highest priority.\\
Tr3 & All the trains are at $S_i$ and requesting for same track to depart.\\
\rowcolor{gray}
Tr4 & Reorder the scheduled trains as per priority.\\
Tr5 & Original departure schedule maintained as highest priority train is departing first as per original schedule.\\
\hline \hline
\rowcolor{gray}
\multicolumn{2}{|c|}{Description of colour tokens of $PN3$}\\
\hline \hline
\rowcolor{gray}
Colour Token(ct) & Description\\
\hline \hline
ct1 & P3 generates it to indicate that train $T_j$ has the highest priority and enables transition Tr2.\\
\rowcolor{gray}
ct2 & P5 generates it when reordering in train departure is decided and enables transition Tr4.\\
ct3 & P5 generates it if original ordering in departure schedule of trains are maintained and enables transition Tr5.\\
\hline
\end{tabular}
}
\end{table}

\noindent \emph{\underline{Analysis of $PN3$:}}\\
Table \ref{tab:PN3} presents the description of
the places $P=\{P1, P2, P3, P4, P5, P6\}$ and transitions $Tr=\{Tr1, Tr2, Tr3, Tr4, Tr5\}$ and
the initial marking is $\boldsymbol M_0=[2, 3, 0, 0, 0, 0]$.
\begin{itemize}
 \item Reachability graph analysis:\\
Similarly, as discussed in subsection \ref{description:Analysis},
here initial marking
$M_0$ is the root node as shown in Figure \ref{fig:PN3_seq}. From our
resultant tree it can be proved that : a) the reachability set
is $R(M_0)$ finite, b)
maximum number of tokens that a place can have is 3, so our
$PN3$ is 3-bounded, c) all transitions can be fired, so there are
no dead transitions.

 \item State equation:\\
 The order of the places in the incidence matrix $\boldsymbol A$
is $P = \{P1, P2, P3, P4, P5, P6\}$, denoted by rows 
and the order of the transitions is $Tr=\{Tr1, Tr2, Tr3, Tr4, Tr5\}$, denoted by columns.\\

\[\tiny
\boldsymbol A = 
\begin{bmatrix}
    -1 & 1 & -1 & 0 & 0 \\
    -1 & 0 & 0 & 0 & 0 \\
     1 & -1 & 0 & 0 & 0 \\
     0 & 1 & -1 & 0 & 0 \\
     0 & 0 & 1 & 0 & 0 \\
     0 & 0 & 0 & 1 & 1 \\
    
\end{bmatrix}
\]


Here, marking $\boldsymbol M=[1, 2, 0, 1, 1, 0]$ is reachable from initial marking
$\boldsymbol M_0=[2, 3, 0, 0, 0, 0]$ through the firing sequence $\sigma_1 = Tr1, Tr2, Tr3, Tr4$.\\ 
$M_0 \xrightarrow{Tr1} M_1 \xrightarrow{Tr2} M_2 \xrightarrow{Tr3} M_3 \xrightarrow{Tr4} M_4$.\\

\[\tiny
\begin{bmatrix}
     2 \\
     3 \\
     0 \\
     0 \\
     0 \\
     0
\end{bmatrix}
+
\begin{bmatrix}
    -1 & 1 & -1 & 0 & 0 \\
    -1 & 0 & 0 & 0 & 0 \\
     1 & -1 & 0 & 0 & 0 \\
     0 & 1 & -1 & 0 & 0 \\
     0 & 0 & 1 & 0 & 0 \\
     0 & 0 & 0 & 1 & 1 \\
    
\end{bmatrix}
\times
\begin{bmatrix}
     1 \\
     1 \\
     1 \\
     1 \\
     0 
\end{bmatrix}
=
\begin{bmatrix}
     1 \\
     2 \\
     0 \\
     1 \\
     1 \\
     0 
\end{bmatrix}
\]
\end{itemize}

\subsection{\textbf{Case 3: Extended Impact of Edge and Node deletion}}
\label{Case3}
Train $T_{j}$ is neither waiting at the station $S_{i}$ where disaster happened, i.e. $P_{jik}=0$
nor on the connecting track, i.e. $L_{jil}=0$. But $T_j$ reaches the station $S_{i}$ within $\tau^{B}$.
\subsubsection{Case 3.1 \newline}
  Train $T_{j}$ is at station $S_{i''}$, where $S_{i''}$ is in neighbourhood of $S_{i}$.\\
  i.e.
  \begin{equation}
  \label{eq:Case3.1a}
  P_{ji''k}=1,~ S_{i''} \in S \setminus S_{i} ~and~ i \in [1,n]
  \end{equation}
 
  If any platform is available at the next station and the connecting track is also free, the system checks for the priority of the train $T_{j}$.
  $T_j$ maintains its original schedule iff it has the highest priority while reaching $S_i$.\\
  i.e. if\\
  \begin{equation}
    \label{eq:Case3.1b}
   (P_{jik'}=0) \wedge (L_{jil}=0) \wedge (Prio(T_{j})|_{t={x_{ji}^{AT}}} > Prio(T_{j'})|_{j\neq j', ~j\in [1,~m]})
  \end{equation}
  then,
  \begin{equation}
    \label{eq:Case3.1c}
   {x_{ji}^{AT}} = {o_{ji}^{AT}}
  \end{equation}

\begin{equation}
  \label{eq:Case3.1d}
 L_{jil}=1 ~and ~P_{jik'}|_{t={x_{ji}^{AT}}}=0
\end{equation}

  \begin{equation}
    \label{eq:Case3.1e}
  P_{jik'}|_{t={x_{ji}^{AT}}}=1 ~and ~L_{jil}=0
  \end{equation}
  Here, damaged platform is $k$.\\
  $k' = \{1, 2, \ldots, p\} \setminus \{k\}$
\subsubsection{Case 3.2 \newline}
\begin{figure}[H]
\centering
\subfigure[]{\label{fig:PN4}\includegraphics[scale=0.80]{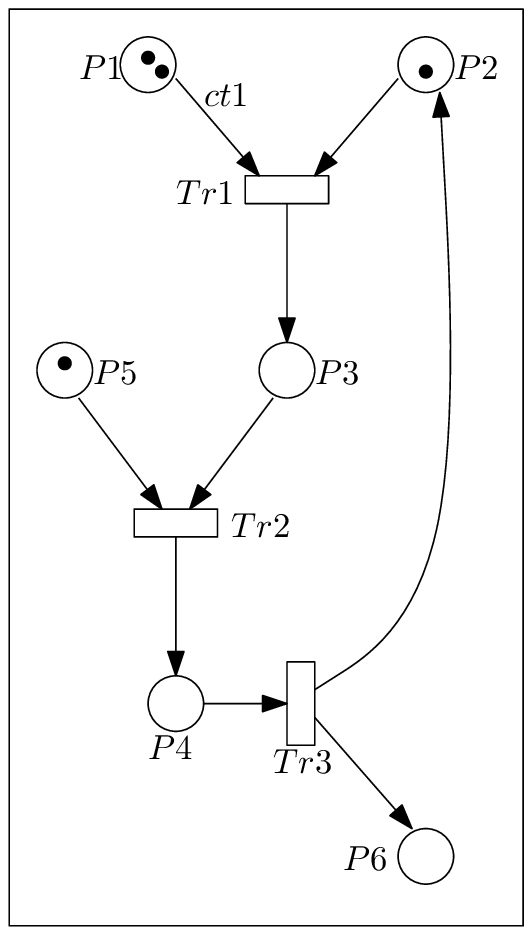}}
\hspace{2.5cm}
\subfigure[]{\label{fig:PN4_seq}\includegraphics[scale=0.750]{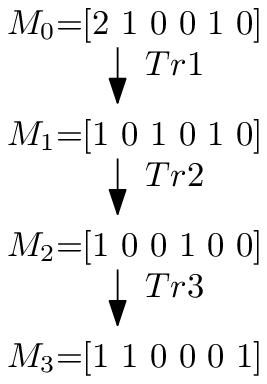}}
\caption{Petri-Net $PN4$
\subcaption{a}{Petri-Net model $PN4$ of Case 3.}
\subcaption{b}{Reachability tree of $PN4$ for different firing sequences.}}
\end{figure}

  Train $T_j$ is at $S_{i'}$, i.e. $P_{ji'k}=1$, where, $i, i' \in [1,n] $ and $i \neq i'$.\\
  There are multiple tracks
  between two stations $S_{i}$ and $S_{i'}$. i.e. $1 < l \leq 4$. If the track $l$ breaks down due to disaster, it is assumed that
  track $l$ is not free. i.e.
  \begin{equation}
    \label{eq:Case3.2a}
   L_{ji'l}=1, ~ 1 \leq l < 4
  \end{equation}
  Then, the trains which are scheduled to use that track face problem.
  In that case, first $S_{i'}$ checks for other available tracks, one of which can be allotted to $T_{j}$,
  provided $T_{j}$ has the highest priority satisfying all the constraints and there is no resource conflict within $\tau^{B}$.
  
  \begin{equation}
    \label{eq:Case3.2b}
   [L_{ji'l'}=0] \wedge [Prio(T_{j})|_{t={x_{ji'}^{DT}}} > Prio(T_{j'})|_{j\neq j', ~j\in [1,~m]}] \wedge [Re(T_{j})|_{x_{ji}^{AT}} \neq Re(T_{j'})|_{x_{ji}^{AT}}]
  \end{equation}
  Figure \ref{fig:PN4} represents Petri-Net model for the scenario of Case 3, described in equations (\ref{eq:Case3.1a})-(\ref{eq:Case3.2b}) and the corresponding
 description of respective places, transitions and tokens are described in Table \ref{tab:PN4}.

\begin{table}[!htb]
\caption{Description of $PN4$}
\label{tab:PN4}

\centering
\renewcommand{\arraystretch}{1.3}
\scalebox{0.70}{
\begin{tabular}{|c||l|}
\hline 
\rowcolor{gray}
\multicolumn{2}{|c|}{Description of Places of $PN4$}\\
\hline \hline
\rowcolor{gray}
Places(P) & Description\\
\hline \hline
P1 & $T_j$ and $T_{j'}$ are at station $S_{i'}$, connecting to station $S_i$ where disaster happens.\\
\rowcolor{gray}
P2 & Any one of the connecting track is free.\\
P3 & Trains are ready to leave.\\
\rowcolor{gray}
P4 & Highest priority train $T_j$ is running on the track.\\
P5 & Platform is free at $S_i$.\\
\rowcolor{gray}
P6 & Highest priority train $T_j$ reaches station $S_i$.\\
\hline \hline
\rowcolor{gray}
\multicolumn{2}{|c|}{Description of Transitions of $PN4$}\\
\hline \hline
\rowcolor{gray}
Transitions(Tr) & Description\\
\hline \hline
Tr1 & It will fire if P1 has more than one tokens and generates ct1 and there is also a token available in P2.\\
\rowcolor{gray}
Tr2 & It will fire if $T_j$ has finished its waiting time at station $S_{i'}$ and a token is available at P5.\\
Tr3 & It will fire if a token is available at P4 indicating $T_j$ is moving forward to $S_i$.\\
\hline \hline
\rowcolor{gray}
\multicolumn{2}{|c|}{Description of colour tokens of $PN4$}\\
\hline \hline
\rowcolor{gray}
Colour Token(ct) & Description\\
\hline \hline
ct1 & P1 generates it to indicate $T_j$ has the highest priority while at $S_{i'}$ and enables transition Tr1.\\
\hline
\end{tabular}
}
\end{table}
\noindent \emph{\underline{Analysis of $PN4$:}}\\
Table \ref{tab:PN4} presents the description of
the places $P=\{P1, P2, P3, P4, P5, P6\}$ and transitions $Tr=\{Tr1, Tr2, Tr3\}$ and
the initial marking is $\boldsymbol M_0=[2, 1, 0, 0, 1, 0]$.
\begin{itemize}
 \item Reachability graph analysis:\\
As discussed in subsection \ref{description:Analysis},
initial marking $M_0$ is the root node as shown in Figure \ref{fig:PN4_seq}. Again,
a) the reachability set
is $R(M_0)$ finite, b)
maximum number of tokens that a place can have is 2, so our
$PN4$ is 2-bounded, c) all transitions can be fired, so there are
no dead transitions.

 \item State equation:\\
Here, in the incidence matrix $\boldsymbol A$,
$ P = \{P1, P2, P3, P4, P5, P6\} $, denoted by rows. 
and the order of the transitions is $Tr=\{Tr1, Tr2, Tr3\}$, denoted by columns.\\
\vspace*{-0.2cm}
\[\small
\boldsymbol A = 
\begin{bmatrix}
    -1 & 0 & 0 \\
    -1 & 0 & 1 \\
    1 & -1 & 0 \\
    0 & 1 & -1 \\
    0 & -1 & 0 \\
    0 & 0 & 1 \\

\end{bmatrix}
\]
%
In our system, marking $\boldsymbol M=[1, 1, 0, 0, 0, 1]$ is reachable from initial marking
$\boldsymbol M_0=[2, 1, 0, 0, 1, 0]$ through the firing sequence $\sigma_1 = Tr1, Tr2, Tr3$.\\ 
$M_0 \xrightarrow{Tr1} M_1 \xrightarrow{Tr2} M_2 \xrightarrow{Tr3} M_3$.
\vspace*{-0.2cm}
\[\small
\begin{bmatrix}
     2 \\
     1 \\
     0 \\
     0 \\
     1 \\
     0
\end{bmatrix}
+
\begin{bmatrix}
    -1 & 0 & 0 \\
    -1 & 0 & 1 \\
    1 & -1 & 0 \\
    0 & 1 & -1 \\
    0 & -1 & 0 \\
    0 & 0 & 1 \\
    
\end{bmatrix}
\times
\begin{bmatrix}
     1 \\
     1 \\
     1 \\
     
\end{bmatrix}
=
\begin{bmatrix}
     1 \\
     1 \\
     0 \\
     0 \\
     0 \\
     1 
\end{bmatrix}
\]
\end{itemize}

\begin{table}[H]
\centering
 \caption{Summary of Disaster Handling Cases Described in Section \ref{Disaster_Handling_and_Rescheduling_Model}}
 \centering 
 \scalebox{0.70}{
 \begin{tabular}{|c||p{10cm}||p{4.5cm}||p{5.5cm}|}
 \hline 
 \rowcolor{gray}
 Case No. & Description & Decision Variable(s) & Decision Taken \\
 \hline \hline
 1 & Station $S_i$ faces problem and train $T_j$ is on track $l$ & $L_{jil}$ & Reroute or Retime \\
 \hline
 \rowcolor{gray}
 1.1 & Station $S_i$ has free platforms when $T_j$ reaches & $P_{jik}, Prio(T_j), Re(T_{j})$ & Reroute or Retime from station \\
 \hline
 1.1.1 & $Re(T_{j})$ is available after $S_{i}$ & $P_{jik}, L_{jil}, x_{ji}^{DT}, x_{ji}^{AT}$ & Reroute from $S_{i}$ \\
 \hline
 \rowcolor{gray}
 1.1.2 & No alternative route found for $T_{j}$ from $S_{i}$ & $x_{ji}^{DT}$ & Delay at $S_{i}$ \\
 \hline
 1.2 & Station $S_i$ has no free platforms when $T_j$ reaches & $P_{jik}, x_{ji}^{AT}$ & Stop on track $l$, Retime \\
 \hline
 \rowcolor{gray}
 2 & Number of trains are about to depart from $S_i$ within buffer time $\tau^B$ & $L_{jil}, P_{jik}, Prio(T_{j}), x_{ji}^{DT}$ & Reorder \\
 \hline
 3 & Train $T_j$ neither waits at affected station $S_i$ nor on track $l$ connected to $S_i$, but reaches to $S_i$ within $\tau^B$ & $L_{jil}, P_{jik}, Prio(T_{j}), x_{ji}^{AT}$ & Retime\\ 
 \hline 
  
 \end{tabular}
 }
 \end{table}
 \subsection{\textbf{Delay Handling}}
 Delay Minimisation can be formulated as:\\
\indent $\bullet$ Delay minimisation at station $S_i$ $(\delta_{ji}^{min})$.\\
\indent $\bullet$ Delay minimisation on the track $l$ $(\delta_{jl}^{min})$.\\
 \noindent $\boldsymbol{\delta_{ji}^{min}}$:
 This aims to minimise the delay in such a way that
 even if the train $T_j$ comes late, it should try to minimise the deviation from scheduled departure time,\\
 i.e.
 \begin{equation}
 \label{eq:Delay_at_station_1}
 x_{ji}^{AT} \geq o_{ji}^{AT} 
 \end{equation}
 \begin{equation}
 \label{eq:Delay_at_station_2}
  x_{ji}^{DT} = o_{ji}^{DT}
 \end{equation}
 So, it compromises dwell time of $T_j$ at $S_i$,\\
 i.e.
 \begin{equation}
 \label{eq:Delay_at_station_3}
 x_{ji}^d < o_{ji}^d 
 \end{equation}
 \noindent $\boldsymbol{\delta_{jl}^{min}}$:
 This aims to minimise the delay considering the journey time (from source to
 destination),\\
 i. e.
 \begin{equation}
 \label{eq:Delay_on_track}
  x_{jl}^J = o_{jl}^J
 \end{equation}
 \subsubsection{\textbf{Evaluating Optimised Objective Function \newline}}
The proposed rescheduling approach aims to minimise the total delay of trains in case of any disaster
while rescheduling.
\indent Objective Function:
\begin{equation}
\label{eq:Objective_function}
 min[\sum_{j} \delta_{j}] = min[\sum_{Rou(T_j)} (\delta_{ji}^{min} + \delta_{jl}^{min})] = min[\sum_{P_{jik}} \delta_{ji}^{min}] + min[\sum_{L_{jil}} \delta_{jl}^{min}]
\end{equation}

\section{Simulation Results}
\label{Experiments_results}
To evaluate the performance of the proposed approach, experiments are conducted
in different scenarios with different combination of tracks,
trains and stations. Both the delay on track and at station are considered in the experiments.
\begin{figure}[H]
\centering
\includegraphics[scale=.350]{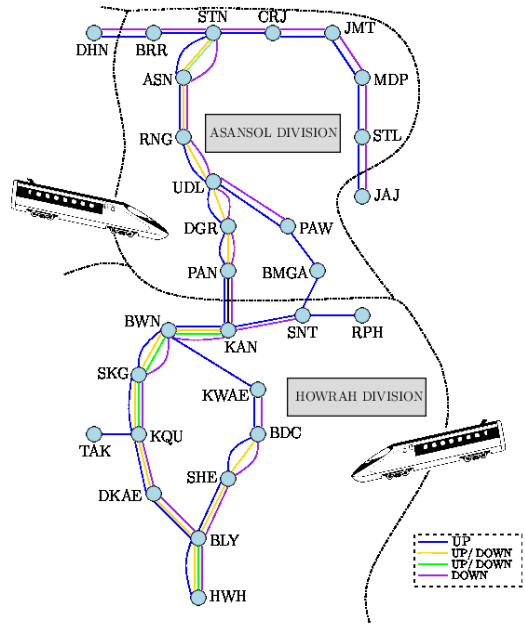}
\caption{Major part of Howrah and Asansol Division, Eastern Railway, India.}
\label{fig:eastern_rail}
\end{figure}

\begin{table}[H]
\caption{Parameter used in Experimental Studies with Station and Train details}
\label{tab:parameter_used}

\centering
\renewcommand{\arraystretch}{1.3}
\scalebox{0.70}{
\begin{tabular}{| p{2cm} || p{10cm} || p{2cm} || p{4cm} | }
\hline 
\rowcolor{gray}
\multicolumn{4}{|c|}{Parameter used}\\
\hline \hline
\rowcolor{gray}
\multicolumn{2}{|  c ||} {Parameter} & \multicolumn{2}{|c|} {Values} \\ 
  \hline \hline
  \multicolumn{2}{|  l ||} {Total number of stations}  & \multicolumn{2}{|c|} {28} \\
   \hline
  \rowcolor{gray}
 \multicolumn{2}{|  l ||} {Max number of platforms at station}  & \multicolumn{2}{|c|} {6}\\
  \hline  
  \multicolumn{2}{|  l ||} {Min number of tracks at station}  & \multicolumn{2}{|c|} {1}\\
  \hline 
   \rowcolor{gray}
  \multicolumn{2}{|  l ||} {Max number of tracks at station}  & \multicolumn{2}{|c|} {4} \\
  \hline
  \multicolumn{2}{|  l ||} {Total number of trains} & \multicolumn{2}{|c|} {21}\\
  \hline
   \rowcolor{gray}
 \multicolumn{2}{|  l ||} {Threshold delay} & \multicolumn{2}{|c|} {30 (in minute)}\\

\hline \hline
\rowcolor{gray}
\multicolumn{4}{|c|}{Station Details of Eastern Railway(Howrah and Asansol division)}\\
\hline \hline
\rowcolor{gray}
Station Code & Station Name & Station Code & Station Name\\
\hline \hline
HWH & Howrah & BMGA & Bhimgara\\
  \hline
  \rowcolor{gray}
  BLY & Bally & PAN & Panagarh\\
  \hline
  SHE & Sheoraphuli Junction & PAW & Pandabeswar\\
  \hline
  \rowcolor{gray}
  DKAE & Dankuni & DGR & Durgapur\\
  \hline
  BDC & Bandel Junction & UDL & Andal\\
  \hline
  \rowcolor{gray}
  KQU & Kamarkundu & RNG & Raniganj\\
  \hline
  TAK & Tarkeshwar &  ASN & Asansol\\
  \hline
  \rowcolor{gray}
  KWAE & Katwa Junction & STN & Sitarampur\\
  \hline
  SKG & Saktigarh & CRJ & Chittaranjan\\
  \hline
  \rowcolor{gray}
  BWN & Barddhaman & JMT & Jamtara\\
  \hline
  KAN & Khana Junction & MDP & Madhupur\\
  \hline
  \rowcolor{gray}
  SNT & Sainthia & STL & Simultala\\
  \hline
  RPH & Rampurhat & JAJ & Jhajha\\
  \hline
  \rowcolor{gray}
  DHN & Dhanbad & BRR & Barakar\\
\hline \hline
\rowcolor{gray}
\multicolumn{4}{|c|}{Train Details of Eastern Railway (Howrah and Asansol division)}\\
\hline \hline
\rowcolor{gray}
 Train No. & Train Name & Category & Priority \\ 
  \hline \hline
  12313 & Sealdah-New Delhi Rajdhani Express & $T_{j}^{Pr}$ & $y_{1}$ \\
  \hline 
  \rowcolor{gray}
  12301 & Howrah - New Delhi Rajdhani Express & $T_{j}^{Pr}$ & $y_{1}$ \\
  \hline 
  12273 & Howrah - New Delhi Duronto Express & $T_{j}^{Pr}$ & $y_{1}$ \\
  \hline 
  \rowcolor{gray}
  12303 & Poorva Express & $T_{j}^{M}$ & $y{2}$ \\
  \hline 
  12019 & Howrah-Ranchi Shatabdi Express & $T_{j}^{Pr}$ & $y_{1}$ \\
  \hline
  \rowcolor{gray}
  22387 & Black Diamond Express & $T_{j}^{P}$ & $y_{4}$ \\
  \hline 
  13051 & Hool Express & $T_{j}^{P}$ & $y_{4}$ \\
  \hline 
  \rowcolor{gray}
  12329 & West Bengal Sampark Kranti Express & $T_{j}^{M}$ & $y_{2}$ \\
  \hline 
  12339 & Coalfield Express & $T_{j}^{P}$ & $y_{4}$ \\
  \hline 
  \rowcolor{gray}
  12341 & Agnibina Express & $T_{j}^{P}$ & $y_{4}$ \\
  \hline 
  13009 & Doon Express & $T_{j}^{M}$ & $y_{2}$ \\
  \hline 
  \rowcolor{gray}
  37211 & Howrah-Bandel Jn Local & $T_{j}^{Lo}$ & $y_{5}$ \\
  \hline 
  13017 & Ganadevta Express & $T_{j}^{P}$ & $y_{4}$ \\
  \hline 
  \rowcolor{gray}
  37911 & Howrah-Katwa Jn Local & $T_{j}^{Lo}$ & $y_{5}$ \\
  \hline 
  63541 & Asansol-Gomoh MEMU & $T_{j}^{Lo}$ & $y_{5}$ \\
  \hline 
  \rowcolor{gray}
  53061 & Barddhaman Jn-Hatia Passenger & $T_{j}^{Lo}$ & $y_{5}$ \\
  \hline 
  15662 & Kamakhya - Ranchi Express & $T_{j}^{M}$ & $y_{2}$ \\
  \hline
  \rowcolor{gray}
  63525 & Barddhaman Jn-Asansol Jn MEMU & $T_{j}^{Lo}$ & $y_{5}$ \\
  \hline 
  63523 & Barddhaman Jn-Asansol Jn MEMU & $T_{j}^{Lo}$ & $y_{5}$ \\
  \hline 
  \rowcolor{gray}
  53131 & Sealdah-Muzaffarpur Fast Passenger & $T_{j}^{P}$ & $y_{4}$ \\
  \hline 
  12359 & Kolkata - Patna Garib Rath Express & $T_{j}^{Pr}$ & $y_{1}$ \\
  \hline
\end{tabular}
}
\end{table}

\subsection{Experimental Setup}
\label{Setup}
The simulation is coded in Java in JADE \cite{jade} in UNIX
platform of personal computer with 2.90 GHz processor speed and 4GB memory.
The results and computations are evaluated under same running environment.
A part of \textit{Eastern Railway, India} \cite{ER},
shown in Figure \ref{fig:eastern_rail},
is taken for experimental studies.\\
\indent
Table \ref{tab:parameter_used} describes the parameters with their values taken for the 
experiments, station details and the train categorisation as discussed in subsection \ref{Classification}. Total 28 major
stations and 21 different types of trains are taken to generate the real-time scenarios.\\
\indent
First the database is set with all the station details and train details and the neighbourhood of the stations 
in railway network. For both the Asansol and Howrah division, there are maximum $4$ tracks in between two stations
and for some part, stations are connected with $1$ single track. Each station is assumed to have max $p=6$ number of platforms.
The permissible threshold delay, as discussed before, is taken as $30$ min. For simplicity we assume, all stations are
equidistant and trains are running at a constant speed throughout its journey.

 \subsection{Illustration}
 \label{Illustration}
To illustrate the proposed method $7$ different
scenarios are taken at different times of a day, based upon the cases discussed in Section \ref{Disaster_Handling_and_Rescheduling_Model}.
Some scenarios describe affected stations, whereas some describe the disruption on track as well as blocked stations.
The first scenario $Sc^1$ describes the blocked station $BLY$, but no track is blocked. The disaster happened at $6:00$ am.
Based on the proposed approach, train no. $12019, 13051, 12303$ are delayed and rescheduled. Total delay encountered here is $15$ min.
Similar incidents are described in $Sc^2$, $Sc^3$, and $Sc^5$ in different times, where only stations are blocked.
Scenario $Sc^7$ highlights a disaster, happened at $21:00$, where track no. $2$, i.e., DOWN track between stations $UDL$ and $RNG$ is blocked.
Train no. $12341, 13009, 12359$ are rescheduled in this case with our proposed approach, facing a total delay of $24$ min. Similar 
scenarios are mentioned in $Sc^4$, $Sc^6$. The details of blocked stations, blocked tracks, affected trains, and 
total delay observed in each scenario is described in Table \ref{tab:scenario_experiment}.\\

 \begin{figure}[!htbp]
\centering
\subfigure[]{\label{fig:graph2}\includegraphics[scale=.550]{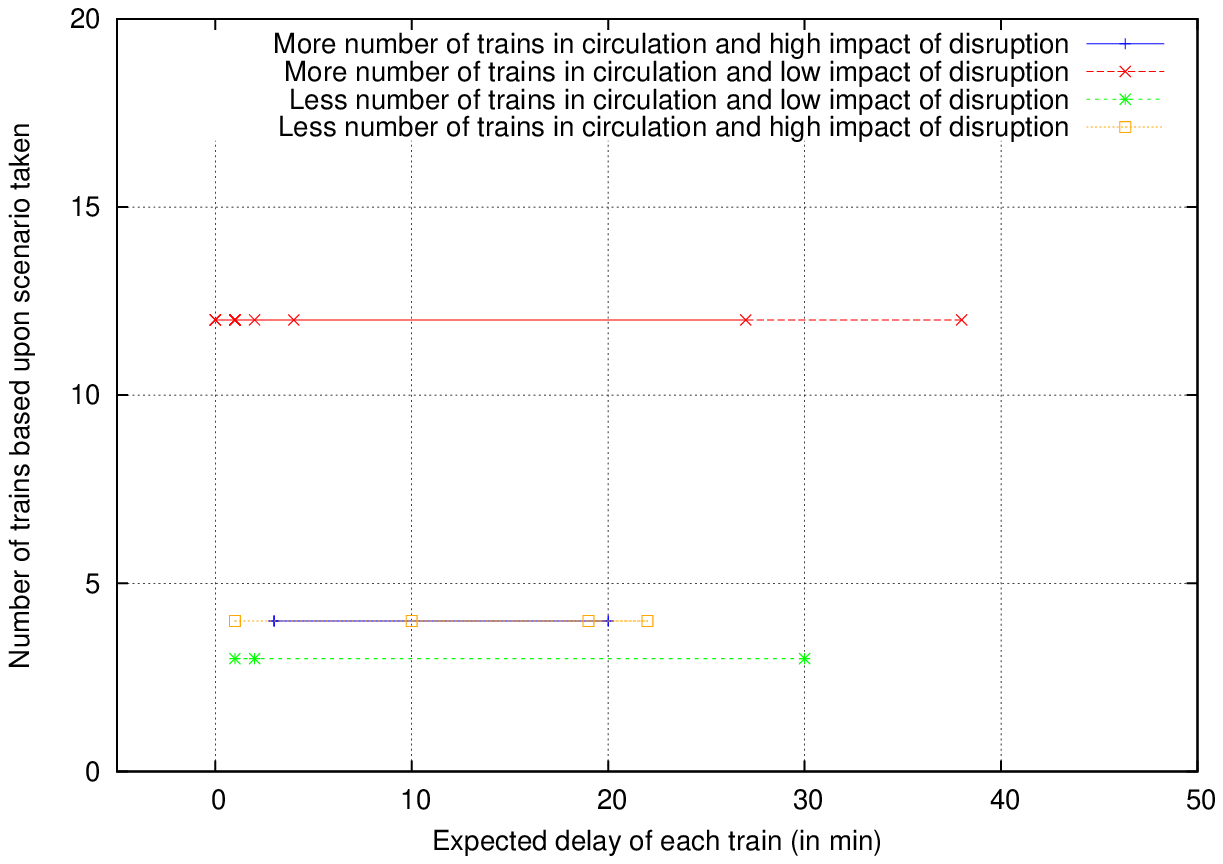}}
\hspace{1cm}
\subfigure[]{\label{fig:graph4}\includegraphics[scale=.550]{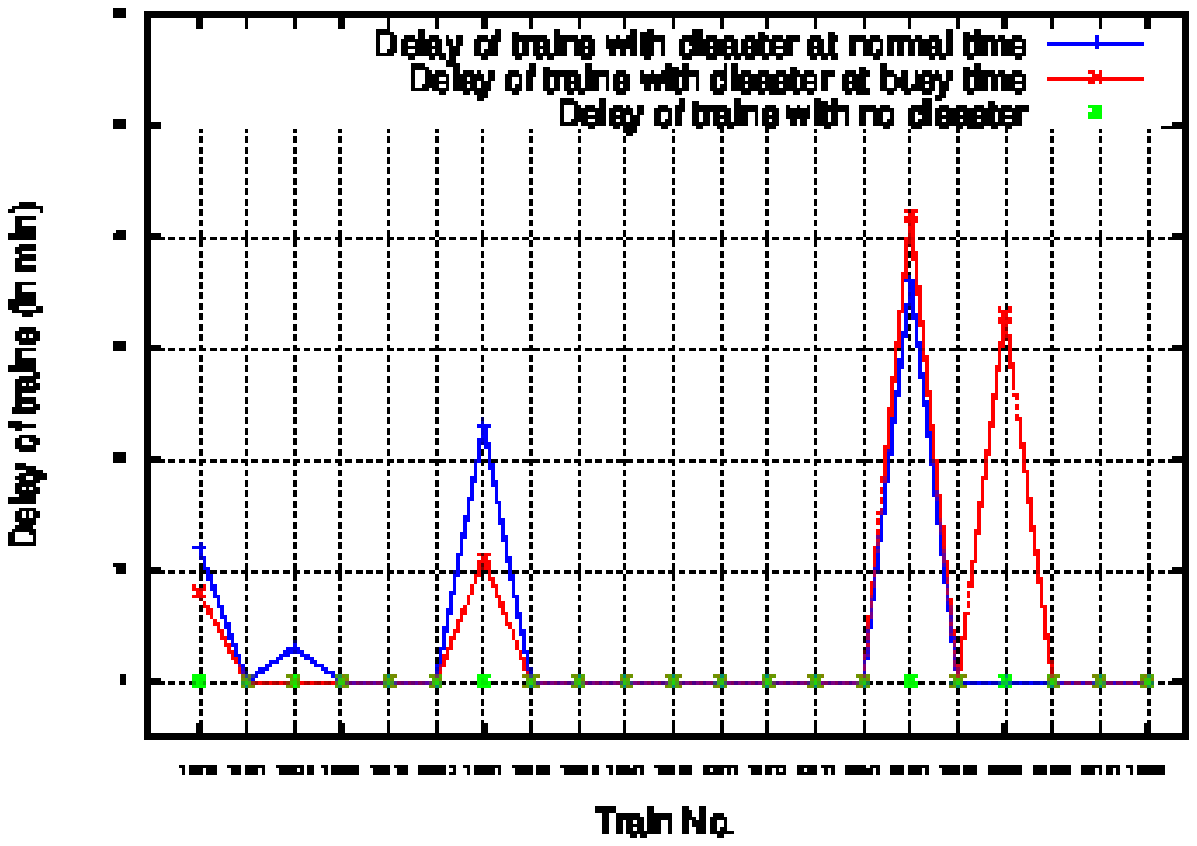}}
\subfigure[]{\label{fig:graph1}\includegraphics[scale=.550]{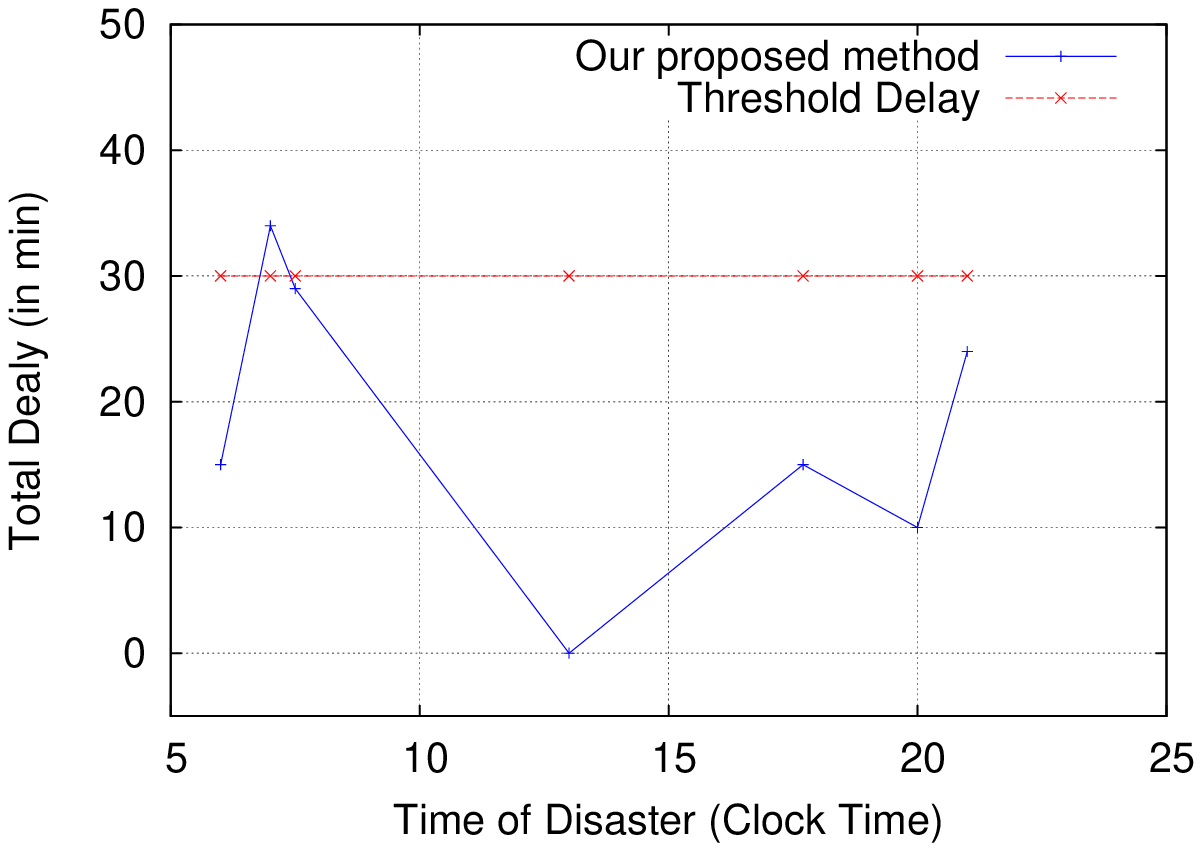}}
\hspace{1cm}
\subfigure[]{\label{fig:graph3}\includegraphics[scale=.550]{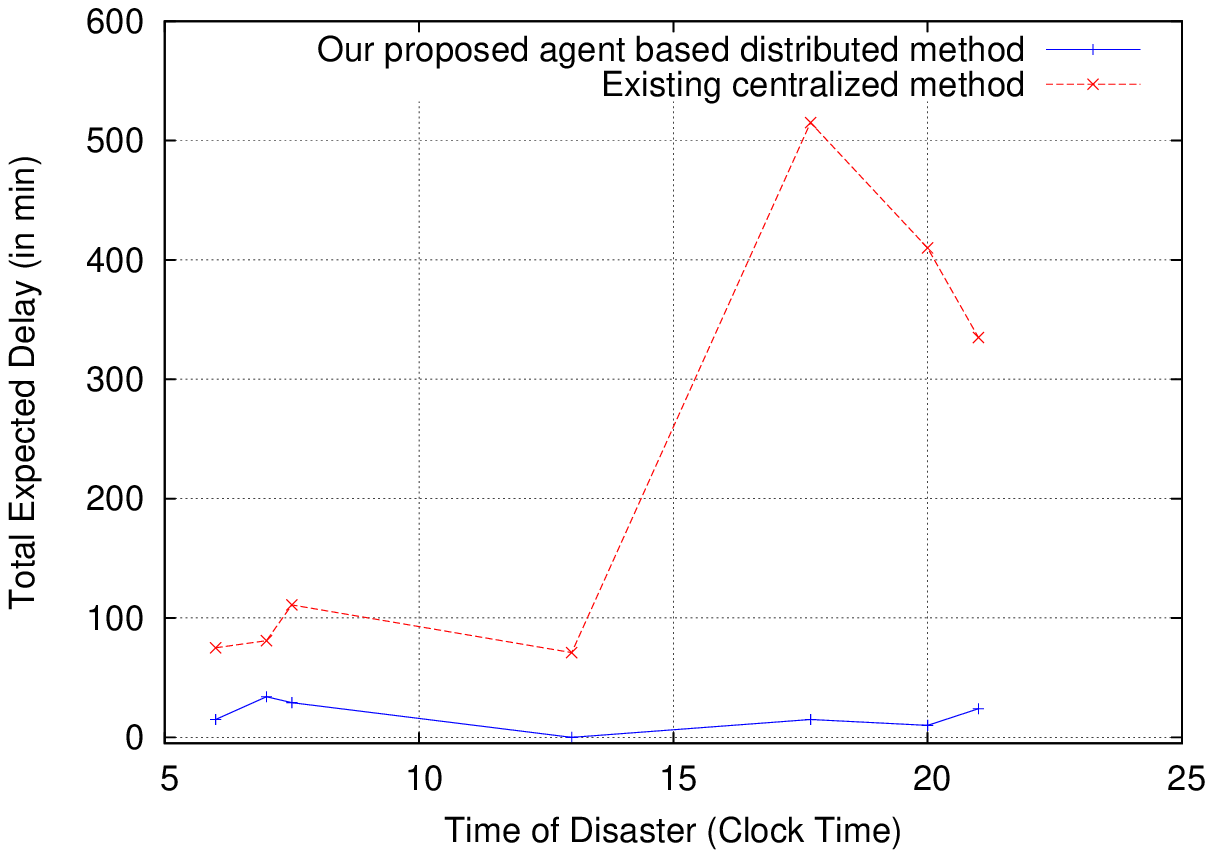}}
\caption{Simulation Results
\subcaption{a}{Change of number of affected trains and their expected delay under disruption scenarios.}
\subcaption{b}{Delay of trains with disaster at normal time and busy time.}
\subcaption{c}{Delay Minimisation through our proposed method.}
\subcaption{d}{Comparison between existing centralised approach and our proposed agent-based distributed approach.}}
\end{figure}

\begin{table}[H]
 \centering
 \caption{Scenarios taken in Experimental Studies}
 \label{tab:scenario_experiment}
 \centering 
  \scalebox{0.70}{
 \begin{tabular}{| c || c || l || l || l || c |}
  \hline 
  \rowcolor{gray}
  	  
  Scenario & Time of disaster & Blocked Station & Blocked Track & Affected Train No. & Total Delay \newline (in min)\\ 
  \hline \hline
  $Sc^1$ & 6:00 & BLY & - & 12019, 13051, 12303 & 15 \\
  \hline
  \rowcolor{gray}
  $Sc^2$ & 7:00 & KAN & - & 12303, 12019, 13051, 53131 & 34 \\
  \hline 
  $Sc^3$ & 7:30 & UDL & - & 12019, 53061, 22387  & 29 \\
  \hline
  \rowcolor{gray}
  $Sc^4$ & 13:00 & BWN, KAN & $(BWN \longleftrightarrow KAN)^2$ & 12273 & 0 \\
  \hline
  $Sc^5$ & 17:40 & KAN & - & 12339, 12313, 12301, 63523, 63525 & 15 \\
  \hline 
  \rowcolor{gray}
  $Sc^6$ & 20:00 & ASN, STN & $(ASN \longleftrightarrow STN)^1$ & 12339, 12341, 12359, 63525 & 10 \\
  \hline
  $Sc^7$ & 21:00 & UDL, RNG & $(UDL \longleftrightarrow RNG)^2$ & 12341, 13009, 12359 & 24 \\
  \hline

 \end{tabular}
  }
 \end{table}
\indent
Depending upon the impact of the occurred disaster, the scenario
is distinguished in four different categories, such as: 
\emph{(a)} More number of trains in circulation and high impact of disruption, \emph{(b)} More number of trains in
circulation and low impact of disruption,
\emph{(c)} Less number of trains in circulation and low impact of disruption, \emph{(d)} Less number of trains in circulation
and high impact of disruption. In figure \ref{fig:graph2}, graphical representation of the expected
delay of each train is shown depending upon this categorisation.\\
\indent 
In Figure \ref{fig:graph4},
the changes in delay of each train is shown
when same disaster happens at normal time and at busy time. Without any disaster, all trains maintain
original schedule and no delay is observed. So, the graph maintains a straight line with \emph{zero} delay for all trains.\\
\indent
Figure \ref{fig:graph1} represents the delay minimisation, achieved through proposed approach.
Threshold delay is taken as $30$ min. We vary the time of disaster in 24h time period to observe the total delay of trains
through the proposed method.\\
\indent
To exhibit the advantage of proposed approach, results are compared with the existing
centralised decision making approach of Indian Railway.
In this method, all the rescheduling decisions are taken by the central authority. 
All the low level authorities pass the necessary messages to the next higher level authority in the railway 
hierarchy and so on. The higher authority checks for all feasible solutions and the best decision message for rescheduling
is passed from central authority to the lower authorities for necessary changes.
This procedure is time consuming and may face disadvantages of traditional centralised systems like, single-point
failure, lesser autonomy, under utilisation etc.
The comparison between existing centralised and the proposed agent-based distributed approach 
is shown in Figure \ref{fig:graph3}. In case of every disaster scenario, happened in different time of a day in the railway network, 
significant reduction in delay is observed through the proposed approach.

\section{Conclusion}
 \label{Conclusion}

This paper proposes a new train rescheduling approach to handle delay
optimisation in case of disruptions in a railway network. An agent based
solution using the DCOP and MDP was designed to address the distributed
nature of the scenario and the uncertainty of disaster recovery time.
Experimental studies are conducted on Eastern Railway, India to evaluate
the effectiveness of the approach. In disastrous situation with noticeably
large recovery time, the proposed approach is shown to produce lower delay
than existing approaches. \\
\indent One of the future research directions will aim
at extending this approach for rescheduling of trains which
follow a meet-pass sequence \cite{110} using headway time \cite{headway}.
This will increase the number of constraints noticeably which need
to be handled efficiently. Further, cross-over
points between any two stations will also be considered which can help in handling various collision scenarios.
\bibliographystyle{unsrt}
\bibliography{iet}

\end{document}